\documentclass[aps, prl, 10pt, twocolumn, a4paper, longbibliography, notitlepage, superscriptaddress]{revtex4-2}

\usepackage{amsmath}
\usepackage{amssymb}
\usepackage{textcomp}
\usepackage{upgreek}
\usepackage{units}
\usepackage{dsfont}
\usepackage{braket} 
\usepackage{hyperref}
\hypersetup{colorlinks=true, citecolor=blue, urlcolor=blue, linkcolor=blue}

\usepackage[pdftex]{graphicx}

\usepackage{microtype}

\usepackage{txfonts}

\renewcommand{\vec}[1]{\ensuremath{\boldsymbol{#1}}}

\begin{document}

\title{
Topological Magnon-Plasmon Hybrids
}

\begin{abstract}
    We study magnon-plasmon coupling in effectively two-dimensional stacks of van der Waals layers in the context of the band structure topology. Invoking the quasiparticle approximation, we show that the magnetic dipole coupling between the plasmons in a metallic layer and the magnons in a neighboring magnetic layer gives rise to a Berry curvature. As a result, the hybrid quasiparticles acquire an anomalous velocity, leading to intrinsic anomalous thermal Hall and spin-Nernst effects in ferromagnets and antiferromagnets. We propose magnetic layers supporting skyrmion crystals as a platform to realize chiral magnon-plasmon edge states, inviting the notion of topological magnon-plasmonics. 
\end{abstract}

\author{Tomoki Hirosawa}
    \email{hirosawa@phys.aoyama.ac.jp}
	\affiliation{Department of Physical Sciences, Aoyama Gakuin University, Sagamihara, Kanagawa 252-5258, Japan}
    
\author{Pieter M.~Gunnink}
	\affiliation{Institute of Physics, Johannes Gutenberg University Mainz, Staudingerweg 7, Mainz 55128, Germany}

\author{Alexander Mook}
	\affiliation{University of Münster, Institute of Solid State Theory, 48149 Münster, Germany}
	
\date{\today}

\maketitle

\paragraph{Introduction.}
The interplay of collective excitations in condensed matter---such as phonons, magnons, and plasmons---is a topic of interest to both fundamental science and technology. The coupling of quasiparticles leads to their hybridization \cite{Kittel1958MagnonPhononCoupling, RahimiIman2020}, resulting in spectral anticrossings that are explored in the context of quantum hybrid systems \cite{BookCavityOptomechanics2014, yuanQuantumMagnonicsWhen2022}, band structure quantum geometry \cite{karzigTopologicalPolaritons2015, LiuTopoPhononics2019, YuReviewNonHermitianTopoMagnons2024, Takahashi2016}, and light-matter coupling \cite{RahimiIman2020}. The coupling of magnons to plasmons, in particular, may serve to control hybrid states of charge density and magnetization fluctuations, thereby interfacing the charge and spin domains in the solid state. While in three dimensions there is a large energy mismatch between magnons and the gapped plasmons in bulk \cite{bar1966interaction, Baskaran1973}, magnon-plasmon coupling in surface magnon polaritons leads to the sought-after anticrossings \cite{Hartstein1973, Matsuura1983, uchidaGenerationSpinCurrents2015, Macedo2019, To2022, Dantas2023}. The discovery of two-dimensional materials that can support gapless plasmons, such as graphene \cite{Hwang2010, Liu2010plasphon, Brar2014}, and magnetic order \cite{Gong2017vdW, Huang2017, Wang2022vdWgenome}, has opened up a promising platform for bulk magnon-plasmon hybridization \cite{costaStronglyCoupledMagnon2023, Ghosh2023, Dyrdal2023,yuan2024,yuan2024strongtunablecouplingantiferromagnetic,GunninkAFMplasmon2025,rudziński2025}. For instance, magnons and plasmons supported in a magnetic monolayer metal have been predicted to hybridize due to spin-orbit coupling \cite{Dyrdal2023,rudziński2025}. Magnon-plasmon hybrids are also proposed to arise from magnetic dipole coupling in a bilayer of graphene and an insulating ferromagnet \cite{costaStronglyCoupledMagnon2023}. 
Excitingly, \emph{topological} spin-plasma waves have been predicted in a heterostructure of a three-dimensional topological insulator and an insulating ferromagnetic layer, where the hybridization is enabled by the strong spin-orbit coupling in the Dirac surface state \cite{Efimkin2021}. Motivated by the existence of topological polaritons (e.g., photon-exciton  \cite{karzigTopologicalPolaritons2015, Li2021TopoExcitonPolariton} or photon-phonon hybrids \cite{Guddala2021}), here we raise the question whether magnon-plasmon coupling in less complex and effectively two-dimensional structures can give rise to band-geometric and topological effects that could lead to signatures in transport, and invite the notion of topological magnon-plasmonics.

\begin{figure}
    \centering
    \includegraphics[width=\columnwidth]{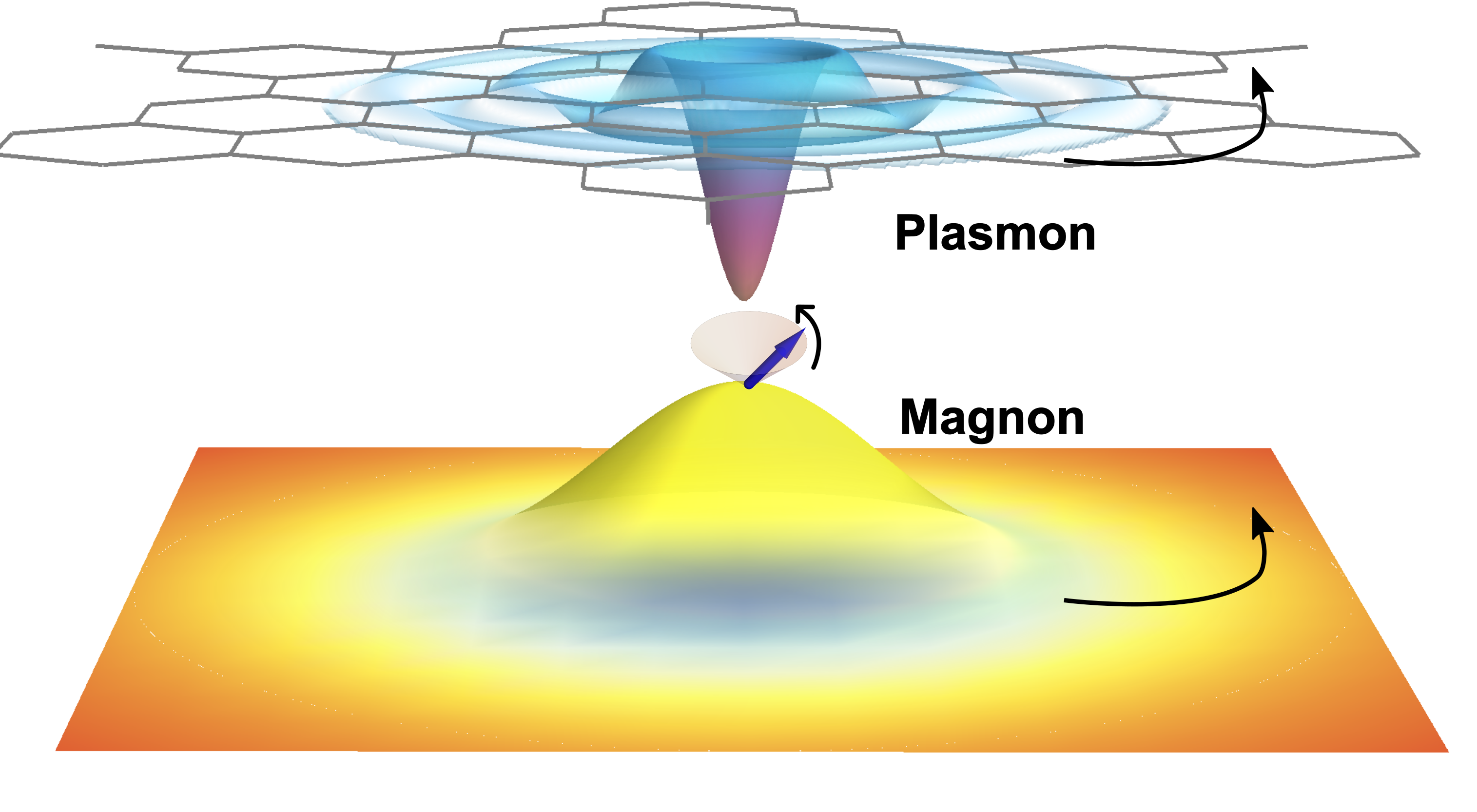}    
    \caption{
     Sketch of topological chiral edge magnon-plasmons in a bilayer of a metal (top layer) and a magnetic system (bottom layer). The magnons are coherently coupled to the plasmons by magnetic dipole coupling, causing a topologically non-trivial spectral anticrossing in the bulk, which supports the chiral hybrid excitations at the edges of the sample (indicated by the water-wave-like feature and the yellow wave packet with an arrow).
     }
    \label{fig:teaser}
\end{figure}

We show that magnon-plasmon coupling in two dimensions gives rise to hybrid quasiparticles with finite Berry curvature, which we calculate explicitly for effective bilayer structures of a metal and a magnetic insulator layer coupled by magnetic dipole coupling. In uniformly ordered magnetic materials, such as ferromagnets and antiferromagnets, we identify finite Chern numbers and anomalous thermal Hall and spin-Nernst transport responses caused by the nontrivial band geometry. 
In addition, in the case of a magnetic skyrmion lattice, schematically shown in Fig.~\ref{fig:teaser}, we study the emergence of chiral magnon-plasmon edge states. These may serve to guide hybrid states of light, charge density oscillations, and magnetic oscillations in magnonic-plasmonic networks and hybridized magnonic materials \cite{To2024}.

\textit{Model.}
Inspired by the setup in Ref.~\onlinecite{costaStronglyCoupledMagnon2023}, we consider a heterostructure of a metallic and a magnetic layer in the $xy$ plane (decoupled by an intermediate spacer). The Hamiltonian of the full system is given by
\begin{align}
    H^{(i)} = H_\text{magnet}^{(i)} + H_\text{plasmon} + H_\text{coupling}.
    \label{eq:full-model}
\end{align}
The effective spin Hamiltonian $H^{(i)}_\text{magnet}$ describes the magnetic system in terms of localized spin operators $\vec{S}_i$, with site index $i$. We provide the details below, where we will consider three cases of increasing complexity: ferromagnets ($i=1$), antiferromagnets ($i=2$), and skyrmion crystals ($i=3$).
The metallic layer is assumed to support plasmons with energy $\Omega_{\vec{k}}$, created (destroyed) by $a^\dagger_{\vec{k}}$ ($ a_{\vec{k}}$), and modeled as noninteracting bosons, so that $H_\text{plasmon} = \hbar\sum_{\vec{k}} \Omega_{\vec{k}} a^\dagger_{\vec{k}} a_{\vec{k}}$. For concreteness, we adopt parameters of graphene to estimate $\Omega_{\vec{k}}$ \cite{ferreiraQuantizationGraphenePlasmons2020}, as reviewed in the Supplemental Material (SM) \cite{SM} (including Refs.~\cite{laturiaDielectricPropertiesHexagonal2018,soykalStrongFieldInteractions2010, Tabuchi2014,Murakami2017,To2023,Shindou13,wilsonChiralSkyrmionsCubic2014,castronetoElectronicPropertiesGraphene2009,Fukui2005}), where we also provide material details of a thin dielectric spacer. 
For the perturbative coupling between the layers, we assume magnetic dipole coupling, $H_\text{coupling} = g \mu_\text{B} \sum_i \vec{B}_\text{p}(\vec{r}_i) \cdot \vec{S}_i$, 
with g-factor $g$, Bohr's magneton $\mu_\text{B}$, and the dynamic magnetic field of the plasmon at coordinate $\vec{r}_i$,
\begin{align}
	\vec{B}_\text{p}(\vec{r}_i)
	&=
	\sum_{\vec{k}}\mathcal{B}_{\vec{k}}(z) 
    \mathrm{e}^{\mathrm{i}\vec{k}\cdot\vec{r}_i} \,\left( \hat{\vec{k}}\times \hat{\vec{z}} \right) \,a_{\vec{k}}+\textrm{h.c.}
    \label{eq:Bfield1}
\end{align}
Its Fourier amplitudes $\mathcal{B}_{\vec{k}}(z)$ have been calculated previously \cite{ferreiraQuantizationGraphenePlasmons2020,costaStronglyCoupledMagnon2023} and are reproduced in the SM \cite{SM}. Note that the momentum reads $\vec{k} = (k_x,k_y,0)^\text{T}$, such that $\vec{B}_\text{p}(\vec{r}_i)$ lies in the $xy$ plane. Hats indicate unit vectors.

\paragraph{Ferromagnet.}
To provide the basic intuition for the topological magnon-plasmon hybridization, we first consider a simple square-lattice ferromagnet, with $N_z$ layers along the $z$-direction, whose spin Hamiltonian reads as 
\begin{align}
    H^{(1)}_\text{magnet}
    =
    - \frac{J}{2} \sum_{\langle i,j \rangle} \vec{S}_i \cdot \vec{S}_j 
    - K \sum_i \left( S_i^z \right)^2,
\end{align}
where $J>0$ is a nearest-neighbor ferromagnetic exchange coupling and $K>0$ an out-of-plane easy-axis anisotropy. After a Holstein-Primakoff transformation about the ferromagnetic ground state, $S^+_i = S^x_i + \mathrm{i} S^y_i \approx \sqrt{2S} b_i$, $S^-_i = (S^+_i)^\dagger$, and $S_i^z = S - b_i^\dagger b_i$ \cite{Holstein1940}, the free magnon Hamiltonian reads
$H^{(1)}_\text{magnet} = \hbar\sum_{\vec{k}} \omega_{\vec{k}} b^\dagger_{\vec{k}} b_{\vec{k}}$, where $b^\dagger_{\vec{k}}$ creates a magnon with momentum $\vec{k}$ and bare energy $ \hbar\omega_{\vec{k}} = 2JS ( 2-\cos k_x - \cos k_y ) + \Delta$, where $\Delta\equiv 2KS$ is the magnon gap. We set the lattice constant to one.

The full bilinear part of $H^{(1)}$ in Eq.~\eqref{eq:full-model} reads
$
    H^{(1)} 
    =
    \frac{1}{2}
    \sum_{\vec{k}}
    \vec{\phi}^\dagger_{\vec{k}}
    H_{\vec{k}}
    \vec{\phi}_{\vec{k}}
$
where
$
    \vec{\phi}^\dagger_{\vec{k}} =
    (
        b^\dagger_{\vec{k}}, a^\dagger_{\vec{k}}, b_{-\vec{k}}, a_{-\vec{k}}
    )
$.
We drop the anomalous couplings ($a^\dagger b^\dagger$ and $a b$) between magnons and plasmons, because their coupling strength is much smaller than the energy at which the bands cross, and thus they give negligible corrections to the hybridization. Thus, the dynamics of the coupled system is approximated by 
$
    H^{(1)} 
    \approx
    \sum_{\vec{k}}
    \vec{\psi}^\dagger_{\vec{k}}
    \tilde H_{\vec{k}}
    \vec{\psi}_{\vec{k}}
$,
with
$
    \vec{\psi}^\dagger_{\vec{k}} =
    (
        b^\dagger_{\vec{k}}, a^\dagger_{\vec{k}}
    )
$
and
\begin{align}
    \tilde  H_{\vec{k}}
    =
    \begin{pmatrix}
        \hbar\omega_{\vec{k}} & G_{k} \mathrm{e}^{ \mathrm{i}\varphi_{\vec{k}}}\\
        G_{k} \mathrm{e}^{-\mathrm{i}\varphi_{\vec{k}}} & \hbar\Omega_{\vec{k}} 
    \end{pmatrix}
    ,
    \quad
    \tan \varphi_{\vec{k}} =  \frac{\hat{k}_y}{\hat{k}_x}.
    \label{eq:FMkernel}
\end{align}
The coupling $G_{k} \mathrm{e}^{ \mathrm{i}\varphi_{\vec{k}}}$ derives from 
$
    \vec{B}_\text{p}(\vec{r}_i) \cdot \vec{S}_i 
    = 
    B^-_\text{p}(\vec{r}_i) S^+_i + B^+_\text{p}(\vec{r}_i)  S^-_i, 
$
with 
$
    B^\pm_\text{p}(\vec{r}_i) = [B^x_\text{p}(\vec{r}_i)\pm \mathrm{i} B^y_\text{p}(\vec{r}_i)]/2 \propto \mathrm{e}^{ \mathrm{i}\varphi_{\vec{k}}}
$.
We provide the full expression of $G_{k}$ in the SM \cite{SM}, but we note here that the $G_{k}\propto \sqrt{N_z}$, since a single plasmon mode couples to the Kittel-like mode along the $z$-direction, formed by $N_z$ layers.

Diagonalizing $\tilde H_{\vec{k}}$ yields two energies $E_{\vec{k}}^\pm$. As the hybridization happens at long wavelengths, we approximate $\omega_{\vec{k}} \approx \omega_{k}$ and $\Omega_{\vec{k}} \approx \Omega_{k}$, where  $k=|\vec{k}|$. The remaining dependence of $H_{\vec{k}}$ on the direction of $\vec{k}$ comes from the phase factors. Thus, the eigenvectors $\vec{w}^\pm_{\vec{k}}$ of $\tilde H_{\vec{k}}$ 
can be written as 
$
    \vec{w}_{\vec{k}}^\pm = U_{\vec{k}} \vec{v}_k^\pm
$, 
where 
$
    U_{\vec{k}} = \text{Diag}( \mathrm{e}^{\mathrm{i} \varphi_{\vec{k}}/2}, \mathrm{e}^{-\mathrm{i} \varphi_{\vec{k}}/2} )
$ 
and $\vec{v}_k^\pm$ are the eigenvectors of $H_{\vec{k} = (k_x,0)}$ for which $\varphi_{\vec{k}} = 0$.
The Berry curvature reduces to 
\begin{align}
    F^\pm_{\vec{k}} 
    = 
    \mathrm{i}
    \left(
        \frac{\partial\vec{w}_{\vec{k}}^{\pm,\dagger}}{\partial k_x}
        \frac{\partial\vec{w}_{\vec{k}}^\pm}{\partial k_y}
        -
        \frac{\partial \vec{w}_{\vec{k}}^{\pm,\dagger}}{\partial k_y}
        \frac{\partial\vec{w}_{\vec{k}}^\pm}{\partial k_x}
    \right)
    = 
        -\frac{1}{2k} \frac{\partial}{\partial k} \left(
        \vec{v}_{k}^{\pm,\dagger} \sigma_3 \vec{v}_{k}^{\pm} 
        \right) ,
\end{align}
with $\sigma_3 = \text{Diag}(1,-1)$; it is rotationally symmetric, $F^\pm_{\vec{k}} = F^\pm_{k}$. The Chern number is obtained as
\begin{align}
    C^\pm
    = 
    \int \frac{F^\pm_{\vec{k}}}{2\pi} \, \mathrm{d}^2 k
    =
    \frac{
        \left( \vec{v}_{k}^{\pm,\dagger} \sigma_3 \vec{v}_{k}^{\pm} \right)_{k \to \infty}
        -
        \left( \vec{v}_{k}^{\pm,\dagger} \sigma_3 \vec{v}_{k}^{\pm} \right)_{k \to 0}
    }{2}
\end{align}
and we find a topological phase with $C^\pm = \pm 1$ \cite{SM}. 
The above considerations of topology agree with those in Ref.~\onlinecite{Okamoto2020} on classically coupled spin waves and electromagnetic fields.

\begin{figure}
    \centering
    \includegraphics{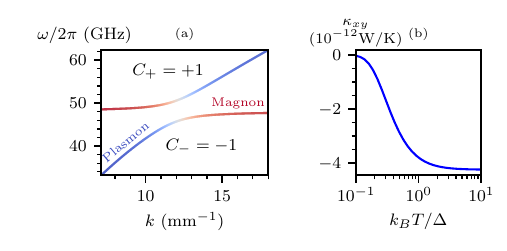}
    \caption{Magnon-plasmon coupling in a ferromagnet-metal bilayer. 
    (a) Quasiparticle dispersion with color indicating magnon and plasmon character. The Chern number $C_n$ of the upper and lower band is indicated.
    (b) Thermal Hall conductivity $\kappa_{xy}$ as a function of temperature $T$. Here $\Delta=\unit[0.2]{meV}$ is the ferromagnetic magnon gap. }
    \label{fig:FM}
\end{figure}

As an example, we choose $S=5/2$, $K=\unit[0.04]{meV}$, $J=\unit[1.15]{meV}$ and $N_z=10^3$ and plot the magnon-plasmon anticrossing in Fig.~\ref{fig:FM}(a), with color indicating the magnon and plasmon character. 
The emerging Berry curvature gives rise to an anomalous velocity and causes an intrinsic thermal Hall effect, that is, a heat current density $q_x = - \kappa_{xy} \partial_y T$ directed along the $x$ direction as response to a temperature gradient $\partial_y T$ in the $y$ direction, with the thermal Hall conductivity $\kappa_{xy}$ given by \cite{Matsumoto2011a, Matsumoto2011b, Matsumoto2014, Qin2011}
\begin{align}
    \kappa_{xy} &= -\frac{k_\text{B}^2T}{\hbar A}
    \sum_{n = \pm}
    \sum_{\vec{k}}
       \left[ c_2(\rho_{n,\vec{k}}) - \frac{\pi^2}{3} \right]F_{\vec{k}}^n ,
\end{align}
where $k_B$ is the Boltzmann constant, $T$ is temperature, and $A$ is the total area of the system.  
Furthermore, $c_{2}(x) = ( 1 + x )  (\ln{\frac{1 + x}{x}})^2 - (\ln{x})^2 - 2 \text{Li}_2(-x)$, where $\text{Li}_2(x)$ is the dilogarithm function, and $\rho_{n,\vec{k}} = [ \mathrm{e}^{E_{n,\vec{k}}/(k_B T)} -1 ]^{-1}$ is the Bose-Einstein function.
As shown in Fig.~\ref{fig:FM}(b), the thermal Hall conductivity gets thermally activated and reaches a plateau around $k_\text{B} T \sim \Delta$. With typical FM resonance frequencies in the range of some tens of GHz, temperatures of a few Kelvin are enough to saturate $\kappa_{xy}$. Here, we have dropped the anomalous couplings in the calculation of the thermal Hall conductivity, but we show in the SM \cite{SM} that including the anomalous couplings does not affect qualitatively the thermal Hall conductivity.

\paragraph{Antiferromagnet.}
As a more complex model, we consider an effectively two-dimensional easy-axis antiferromagnet on the square lattice described by the spin Hamiltonian
\begin{align}
    H_\text{magnet}^{(2)} = \frac{J}{2} \sum_{\langle i,j \rangle } \vec{S}_i \cdot \vec{S}_j - K \sum_i \left( S_i^z \right)^2 ,
    \label{eq:spinHamAFM}
\end{align}
where $J>0$ is antiferromagnetic nearest-neighbor exchange interaction and $K>0$ out-of-plane easy-axis anisotropy.

We perform an antiferromagnetic Holstein-Primakoff transformation \cite{SM}, after which the bilinear Hamiltonian reads
$
    H^{(2)}
    \approx
    \frac{1}{2}
    \sum_{\vec{k}}
    \vec{\Phi}^\dagger_{\vec{k}}
    H_{\vec{k}}
    \vec{\Phi}_{\vec{k}}
$
in the basis
$\vec{\Phi}^\dagger_{\vec{k}} = (b_{\vec{k},1}^\dagger, b_{\vec{k},2}^\dagger,b_{-\vec{k},1}, b_{-\vec{k},2},a^\dagger_{\vec{k}}, a_{-\vec{k}})$, where $b_{\vec{k},n}^\dagger$ creates a spin flip on the $n$th sublattice of the antiferromagnet ($n\in\{1,2\}$). 
The Hamilton matrix $H_{\vec{k}}$ has to be paraunitarily diagonalized \cite{Colpa1978} to find the three physical excitations.

The solution of the coupled magnon-plasmon system can be simplified by first diagonalizing the magnon sector, and then dropping the remaining anomalous magnon-plasmon coupling \cite{SM}. 
The reduced system assumes the form 
$
    H^{(2)}
    \approx
    \sum_{\vec{k}}
    \tilde{\vec{\Phi}}^\dagger_{\vec{k}}
    \tilde{H}_{\vec{k}}
    \tilde{\vec{\Phi}}_{\vec{k}}
$,
where 
$
\tilde{\vec{\Phi}}^\dagger_{\vec{k}} = (\alpha^\dagger_{\vec{k}}, \beta^\dagger_{\vec{k}}, a^\dagger_{\vec{k}})
$
contains the magnon normal mode operators $\alpha^\dagger_{\vec{k}}$ and $\beta^\dagger_{\vec{k}}$ that create a magnon with frequency $\omega_{\vec{k},\pm} = \omega_{\vec{k}}$.
The reduced kernel
\begin{align}
    \tilde{H}_{\vec{k}} = \begin{pmatrix}
        \hbar\omega_{\vec{k}} & 0 & \tilde{G}_{\vec{k}} \\
        0 & \hbar\omega_{\vec{k}} & \tilde{G}^\ast_{\vec{k}} \\
        \tilde{G}^\ast_{\vec{k}} & \tilde{G}_{\vec{k}} & \hbar\Omega_{\vec{k}}
    \end{pmatrix}
    \label{eq:AFMmatrix}
\end{align}
has an arrow-head matrix structure, and $\tilde{G}_{\vec{k}} = f_{k} G_{k} \mathrm{e}^{\mathrm{i}\varphi_{\vec{k}}}$ is the effective magnon-plasmon coupling that---compared to the ferromagnetic case---comes with an enhancement factor $f_{k} > 1$, as already pointed out in Ref.~\cite{Dyrdal2023}, see also SM \cite{SM}.  

\begin{figure}
    \centering
    \includegraphics{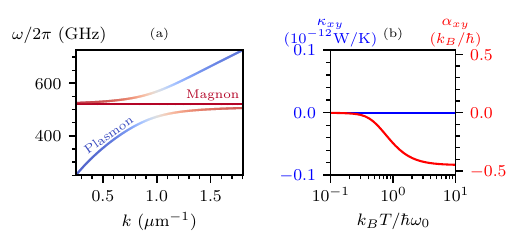}
    \caption{Magnon-plasmon coupling in an antiferromagnet-metal bilayer.
    (a) Quasiparticle dispersion, with color indicating the magnon and plasmon character.
    (b) Thermal Hall conductivity $\kappa_{xy}$ (blue) and spin Nernst conductivity $\alpha_{xy}$ as a function of temperature $T$.  Here  $\hbar\omega_0=\unit[1.5]{meV}$ is the exchange-enhanced antiferromagnetic magnon gap.}
    \label{fig:AFM}
\end{figure}

Given the absence of an external magnetic field, the magnons are spin degenerate and the coupling to the plasmons leads to a spin mixing. We find two anticrossing modes and one (purely magnon-like) mode in the gap [Fig.~\ref{fig:AFM}(a)]. Here we have chosen the same parameter values for $J$, $K$, $S$ and $N_z$ as for the ferromagnet. The Berry curvature is identically zero, because of an effective time-reversal and inversion symmetry. As a result, $\kappa_{xy}$ is absent, see Fig.~\ref{fig:AFM}(b).

Besides heat, magnons carry spin, and a temperature gradient can lead to a transverse spin current, which is quantified by a magnon spin Nernst conductivity \cite{Zyuzin2016SNE, Cheng2016SNE}
\begin{equation}
    \alpha_{xy} = \frac{k_B}{A \hbar} \sum_n \sum_{\vec{k}} c_1(\rho_{n,\vec{k}}) \Omega^{S_z}_{n,\vec{k}} .
\end{equation}
Here, $c_1(x)\equiv x\ln x-(1+x)\ln(1+x)$ and  
\begin{align}
    \Omega^{S_z}_{n,\vec{k}} = i \sum_{m \ne n} \frac{\langle n | j_x^S
 |m\rangle \langle m|\partial_{k_y} \tilde H_{\vec{k}} |n\rangle - \langle n | j_y^S |m\rangle\langle m|\partial_{k_x} \tilde H_{\vec{k}} |n\rangle }{(\varepsilon_{\vec{k},m}-\varepsilon_{\vec{k},n})^2}
\end{align}
is the magnon spin Berry curvature, where we defined the spin current $\vec{j}^S = ( S_z \vec{v} + \vec{v} S_z )/2$ with velocity $\vec{v} = \hbar^{-1} \partial \tilde H_{\vec{k}} / \partial \vec{k}$ and spin $S_z = \hbar\, \text{Diag}(1,-1,0)$.
As shown in Fig.~\ref{fig:AFM}(b), we find $\alpha_{xy}$ is non-zero, which is consistent with time-reversal symmetry. We conclude that the coupling to plasmons can be used to engineer spin currents in simple antiferromagnets. Including the anomalous couplings does not qualitatively affect the magnon spin Nernst conductivity at low temperatures, as shown in the SM \cite{SM}.

\paragraph{Skyrmion crystals.}

In both the ferromagnet and the antiferromagnet, we have found a bulk anticrossing, which in the ferromagnet gives rise to nonzero Chern numbers. However, we do not expect to see well-defined chiral edge states because the bare dispersions, $\omega_{k}$ and $\Omega_{k}$, increase monotonously, resulting in a surface spectrum that does not have a band gap to support edge states. Thus, we next explore magnetic systems that naturally feature magnetically active magnon bands with negative curvature in the long-wavelength limit: magnetic skyrmion crystals (SkX). In particular, we focus on counterclockwise rotation (CCW) mode \cite{mochizukiSpinWaveModesTheir2012} with a negative effective mass.

Topological spin textures have been found in a variety of van der Waals magnets, including metals (Fe$_3$GeTe$_2$~\cite{dingObservationMagneticSkyrmion2020}, WTe$_2$/Fe$_3$GeTe$_2$ heterostructures~\cite{wuNeeltypeSkyrmionWTe22020}, and Cr$_2$Ge$_2$Te$_6$~\cite{hanTopologicalMagneticSpinTextures2019}) as well as insulators (CrBr$_3$~\cite{grebenchukTopologicalSpinTextures2024}).
Furthermore, a large Dzyaloshinskii-Moriya interaction was predicted even in a monolayer of Janus van der Waals magnets~\cite{liangVeryLargeDzyaloshinskiiMoriya2020, xuTopologicalSpinTexture2020, yuanIntrinsicSkyrmionsMonolayer2020}.
Inspired by these experimental observations and computational predictions, we consider skyrmion-hosting van der Waals magnets stacked on graphene and a dielectric spacer.
A minimal spin-lattice model is defined on a square lattice as
\begin{eqnarray}
\label{eq: SpinLatticeH}
    H^{(3)}_\text{magnet}
    &=&-\frac{J}{2}\sum_{\braket{i,j}}\vec{S}_{i}\cdot \vec{S}_{j}+\frac{1}{2}\sum_{\braket{i,j}}\vec{D}_{i,j}\cdot\vec{S}_{i}\times\vec{S}_{j}\nonumber\\
&+&g\mu_{B} B_z \sum_{i} \vec{S}_{i}\cdot\hat{\vec{z}} -K\sum_{i} (S_{i}^z)^2\,,
\end{eqnarray}
with the interfacial Dzyaloshinskii-Moriya interaction $\vec{D}_{i,j}=D\hat{z}\times (\vec{r}_i-\vec{r}_j)/|\vec{r}_i-\vec{r}_j|$.

\begin{figure}
    \centering
    \includegraphics[width=\columnwidth]{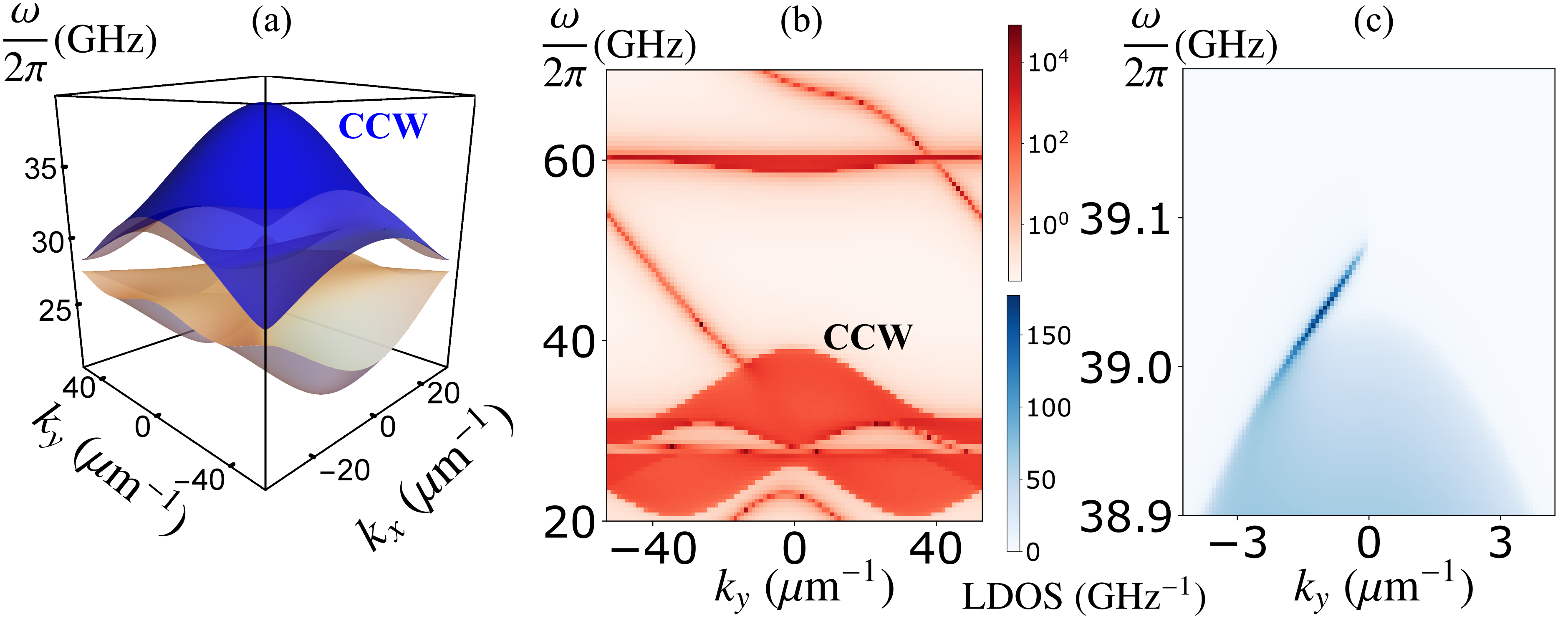}    
    \caption{
    Topological magnon-plasmon polariton in a heterostructure of SkXs and graphene.
    (a) Three-dimensional magnon band structures of the metastable zero-field SkX. The energy and wave vector are rescaled from the dimensionless spin-lattice model as detailed in the SM \cite{SM}. 
    (b,c) Local density of states (LDOS) of magnon-plasmon hybrids at an edge of the semi-infinite skyrmion lattice.  
     }
    \label{skx_band}
\end{figure}

The SkX becomes the ground state due to the competition between ferromagnetic exchange and Dzyaloshinskii-Moriya interactions at finite magnetic fields, but can be stabilized as a metastable state without external fields~\cite{dingObservationMagneticSkyrmion2020,parkEeltypeSkyrmionsTheir2021}.
We obtain the classical spin configuration of SkXs by Monte Carlo annealing, which is relaxed via Landau-Lifshitz-Gilbert simulations.
Using linear spin wave theory~\cite{RoldanMolina2016}, we compute the magnon band structure of the zero-field SkX as shown in Fig.~\ref{skx_band}(a).
We adjust the scaling of the energy and wave vector by assuming $J=10$~meV, $D=1$~meV, $K_z=0.025$~meV, and the size of magnetic unit cell $L=100~$nm~\cite{SM}.

Since the CCW mode gives rise to the rotation of the in-plane magnetization~\cite{hirosawaMagnetoelectricCavityMagnonics2022, SM}, the magnon-plasmon coupling induces a winding in hybridized wave functions with a topological gap.
In addition, the topological spin textures of SkXs support the nontrivial magnon band topology~\cite{Diaz2020FM, Hirosawa2020, hirosawaLasercontrolledRealReciprocalspace2021}.
While the CCW mode is characterized by $C=-1$, the magnon-plasmon coupling induces $C=+1$ below the hybridization gap.
Thus, chiral edge modes of magnons and magnon-plasmon hybrids propagate in opposite directions, which might result in backscattering.
However, a significant mismatch in energy and wave vector of these edge modes suppresses scattering among them.
Consequently, both edge modes should remain robust, although they are not topologically protected.

We construct the tight-binding model of magnon-plasmon hybrids in SkXs by fitting the dispersion of plasmons with quadratic terms~\cite{SM}.
Employing the renormalization technique for the semi-infinite geometry~\cite{HENK199369, Mook2014edge}, we compute the local density of states (LDOS) of magnon-plasmon hybrids at an edge of the semi-infinite skyrmion lattice as shown in Fig.~\ref{skx_band}(b) and (c), respectively. 
The damping rate is set at $\kappa_\textrm{mag}=\kappa_\textrm{pl}=4\textrm{ MHz}\approx 10^{-4}\omega_\textrm{res}$ with the resonance frequency $\omega_\textrm{res}=39$~GHz. Since the coupling strength between the CCW and plasmon modes is estimated as $G\sim 10^{-2}\omega_\textrm{res}$~\cite{SM}, the strong coupling regime is realized in this setup. 
Figure~\ref{skx_band}(b) shows the LDOS over the one-dimensional first Brillouin zone of semi-infinite SkXs.
The magnonic chiral edge mode is found inside the topological gap of the CCW mode, while
it does not exhibit a clear signature of the anticrossing with plasmons.
With a high resolution in energy and wave vector in Fig.~\ref{skx_band}(c), we obtain the magnon-plasmon edge state that connects the lower magnon branch with the upper plasmon branch.
The LDOS of plasmon bands is not visible in Fig.~\ref{skx_band}(c) due to the significant difference in their effective masses. 
The group velocity of the magnon-plasmon edge mode is opposite to the magnonic edge mode above the CCW mode, which is consistent with the Chern number calculation.

We also investigate the effect of damping in the magnon-plasmon edge state~\cite{SM}.
As the damping rate of magnon modes increases, it becomes harder to identify the magnon-plasmon edge state due to the broadening of the bulk spectrum. We find that the chiral edge state is robust up to $\kappa_\textrm{mag}=10^{-3}\omega_\textrm{mag}$. In contrast, the magnon-plasmon edge state is less sensitive to the damping rate of plasmons, indicating the dominant magnonic contributions in the LDOS.
The damping rate of plasmons was recently measured to be $3.1\times 10^{-4}\omega_\textrm{pl}$ in the infrared light~\cite{niFundamentalLimitsGraphene2018}.
In addition, the damping rate of magnons in magnetic insulators is approximated as $\kappa_\textrm{mag}=\alpha \omega_\textrm{mag}$, where the lowest value of the Gilbert damping~$\alpha$ is known as $\alpha=3\times 10^{-5}$ at room temperature in yttrium iron garnet~\cite{10.1063/1.4977423}.
Thus, sufficiently small magnon and plasmon damping rates are attainable at cryogenic conditions.

The hybrid nature of the chiral edge state is advantageous for experimental observations.
Even though observing the chiral edge state has been challenging in topological magnonics, they have been successfully observed in topological plasmonics and photonics~\cite{wuDirectObservationValleypolarized2017, ozawaTopologicalPhotonics2019a}.
For example, the plasmonic chiral edge state can be excited by gold antennas \cite{Low2016} and detected by the near-field microscropy~\cite{wuDirectObservationValleypolarized2017, niFundamentalLimitsGraphene2018} and the transmission spectrum~\cite{wangObservationUnidirectionalBackscatteringimmune2009}. 
In addition, magnonic components can be measured by the NV center magnetometry~\cite{duControlLocalMeasurement2017a, purserSpinwaveDetectionNitrogenvacancy2020} and near-field Brillouin light scattering~\cite{jerschMappingLocalizedSpinwave2010}.

\paragraph{Discussion and conclusion.}
We have identified topologically nontrivial magnon-plasmon hybrids and chiral magnon-plasmon edge states in effectively two-dimensional stacks of van der Waals layers. The resulting nontrivial band geometry was shown to lead to transverse heat and spin transport and to chiral edge magnon-plasmon hybrids. Our effective theory and, in particular, the lattice regularization required to calculate the edge spectrum in Fig.~\ref{skx_band}(b,c), rely on the quasiparticle approximation and neglect the Landau damping in the electron-hole continuum at larger momenta. For open boundaries, the magnon-plasmon gap is filled with a continuum of states. We expect that the hybridization of single-particle states with the continuum will lead to breakdown effects similar to those discussed in Ref.~\onlinecite{Habel2023} for interacting topological magnon insulators and in Ref.~\onlinecite{Hawashin2024} for a generic bosonic toy model. 
However, as long as the energy scale of these effects is smaller than the topological bulk gap, our effective treatment captures the relevant single-particle physics, and an effective non-Hermitian description as in Ref.~\onlinecite{McClarty2019} could be pursued. To avoid the continuum altogether, one could use intrinsically undamped plasmon modes in narrow electron bands, e.g., in Moire graphene bilayers \cite{Lewandowski2019}. 

For the coupling of magnons to plasmons, we have relied on the polaritonic component and magnetic dipole coupling, a small effect but sufficient for a proof of principle. Importantly, magnon-plasmon coupling can occur in a single material with spin-orbit coupling \cite{Dyrdal2023}, which also gives rise to the crucial phase winding necessary for the Berry curvature. 
The range of possible material platforms for the realization of topological magnon plasmons is thus not limited to the stacked layers considered here. The case of plasmons hybridizing with magnon continua, as studied in Refs.~\onlinecite{Ghosh2023, GunninkAFMplasmon2025}, could also be attractive for the realization of chiral spin-plasma edge modes. A critical challenge will be to enhance the magnon-plasmon cooperativity by reducing the quasiparticle damping and increasing the coupling to stabilize wide topological gaps supporting chiral modes, which may find application in nonreciprocal devices such as circulators.

\paragraph{Note added.}
After submission of our work, we became aware of related studies on topological plasmon–magnon modes in magnetic heterostructures of a Rashba electron gas and a magnetic system~\cite{finniganTopologicalHybridisationPlasmons2026}.

\begin{acknowledgments}
\textit{Acknowledgments.} We thank Johannes Knolle, Daniel Malz, Peng Rao, and Ant\'{o}nio Costa for stimulating discussions.
This work was funded by the Deutsche Forschungsgemeinschaft (DFG, German
Research Foundation) - Project No.~504261060. 
T.~H. is supported by JSPS KAKENHI Grant Number JP23K13064 and Aoyama Gakuin University Research Institute “Early Eagle” grant program for promotion of research by early career researchers. P.~G. acknowledges financial support from the Alexander von Humboldt postdoctoral fellowship.
\end{acknowledgments}

\paragraph{Data availability.}
The data that support the findings of this article are openly available~\cite{hirosawa_2025_14890998}.

\end{document}


\title{
Supplementary Material: \\Topological Magnon-Plasmon Hybrids
}
\author{Tomoki Hirosawa}
	\affiliation{Department of Physical Sciences, Aoyama Gakuin University, Sagamihara, Kanagawa 252-5258, Japan}

\author{Pieter M.~Gunnink}
	\affiliation{Institute of Physics, Johannes Gutenberg University Mainz, Staudingerweg 7, Mainz 55128, Germany}

\author{Alexander Mook}
	\affiliation{University of Münster, Institute of Solid State Theory, 48149 Münster, Germany}
	
\date{\today}
\maketitle

\tableofcontents

\section{Graphene Plasmons}
\label{sec: plasmon}
\begin{figure}
    \centering
    \includegraphics[width=100mm]{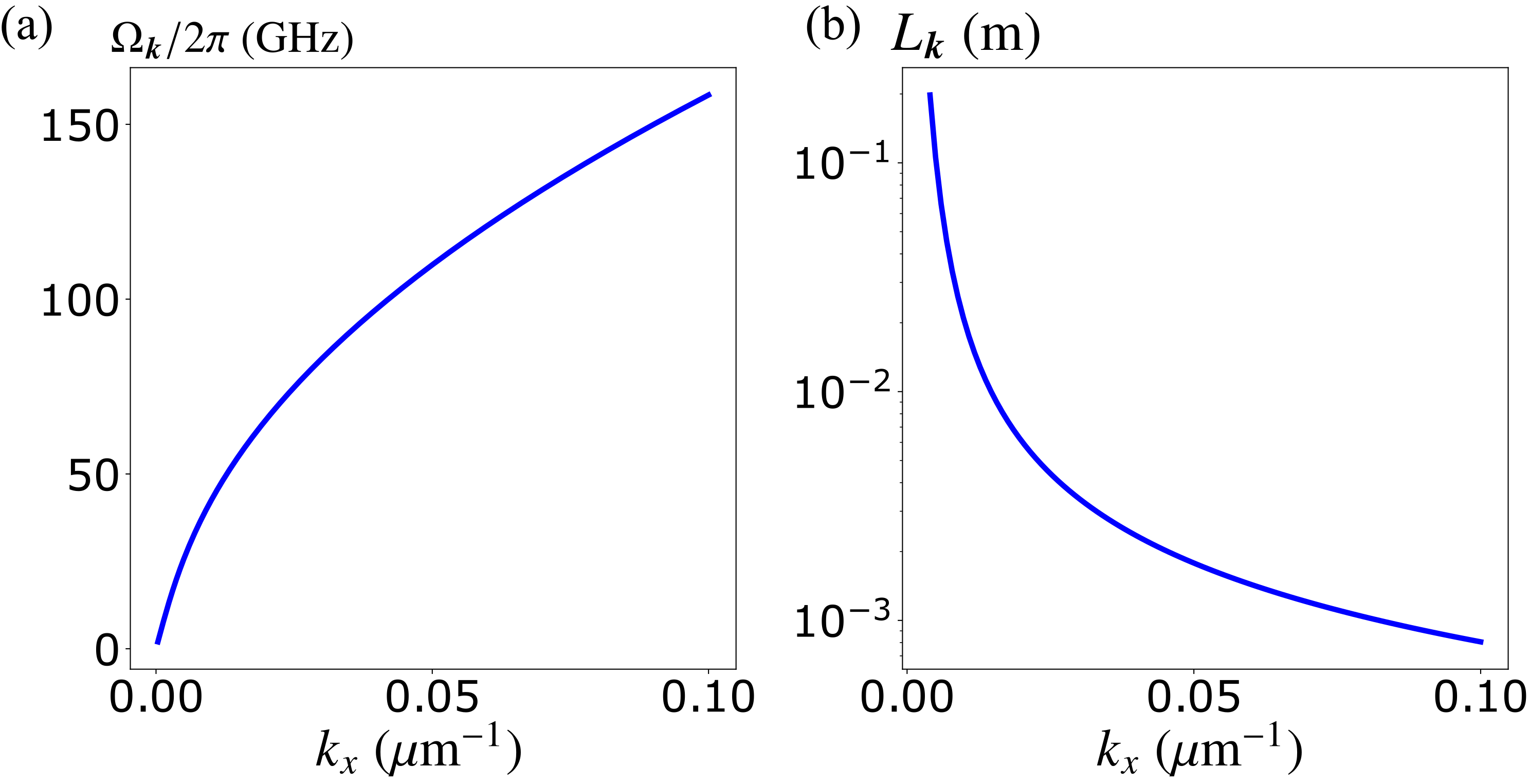}    
    \caption{Dispersion and mode length of plasmon mode in a graphene sheet. Parameters are taken as $v_F=10^6$~m/s, $E_F=100$~meV, $\epsilon_{\parallel}=65$, and $\epsilon_{\perp}=7$.
     }
    \label{graphene_plband}
\end{figure}

We briefly review the coupling between magnons and graphene plasmons~\cite{ferreiraQuantizationGraphenePlasmons2020,costaStronglyCoupledMagnon2023}. 
We consider plasmons in a single layer of graphene at $z=0$, which lies between two dielectric materials.
We assume that the dielectric tensor is anisotropic with 
\begin{equation*}
	\epsilon(z)
	=
	\begin{pmatrix}
	\epsilon_\parallel(z) &0&0\\
	0&\epsilon_\parallel(z) &0\\
	0&0&\epsilon_\perp(z)
	\end{pmatrix}
	.
\end{equation*}
Denoting the dielectric constant $\epsilon_1$ for $z<0$ and $\epsilon_2$ for $z>0$, the magnetic field of the $p$-polarized evanescent waves for $z>0$ is given by~\cite{costaStronglyCoupledMagnon2023}
\begin{align}
	\vec{B}_p(\vec{r})
	&=
	\nabla\times \vec{A}_p(\vec{r},t)
	=
	\sum_{\vec{k}}\mathcal{\vec{B}}_{k}(z)e^{\textrm{i}\vec{k}\cdot\vec{r}} \,(\vec{\hat{k}}\times \hat{z}) \,a_{\vec{k}}+\textrm{h.c.},
 \label{eq:Bfield1}
\end{align}
with 
\begin{align}
	\mathcal{\vec{B}}_{k}(z)
	=
	-\textrm{i}\frac{\epsilon_{2,\parallel} \Omega_{k}}{c\kappa_{2,k}}\sqrt{\frac{\hbar\Omega_{k}\mu_0}{2AL_{k}}}e^{-\kappa_{2,k}z},
  \label{eq:Bfield2}
\end{align}
where $k = |\vec{k}|$, $\kappa_{n,k}=\sqrt{k^2-\frac{\epsilon_{n,\parallel}\Omega^2_{k}}{c^2}}$, $c$ is the velocity of light, and $A$ is the surface area of the graphene sheet. 
Importantly, Eq.~\eqref{eq:Bfield1} highlights that the magnetic field generated by a plasmon has a winding in $\bm k$-space, and can thus provide the nontrivial topology. 
The dispersion relation is approximately given by~\cite{ferreiraQuantizationGraphenePlasmons2020}
\begin{equation}
	 \Omega_{k}^2
	=
	\frac{1}{2}\frac{D}{2\epsilon_0\epsilon_{\parallel}}\sqrt{4k^2+\left(\frac{D}{2\epsilon_0c^2}\right)^2-\frac{4\beta^2k^2\epsilon_{\parallel}}{c^2}}-\frac{1}{2}\left(\frac{D}{2\epsilon_0\sqrt{\epsilon_{\parallel}}c}\right)^2+\beta^2k^2,
\end{equation}
with the average in-plane dielectric constant $\epsilon_{\parallel}$. Assuming the linear dispersion in the graphene band, other parameters are given as $\beta^2=v_F^2/2$ and $D=e^2E_F/(\pi \hbar^2)$ with $v_F$ is the Fermi velocity, and $E_F$ is the Fermi energy. 
The mode length $L_{k}$ is given by~\cite{ferreiraQuantizationGraphenePlasmons2020}
\begin{align}
    L_{k}
    =
    \frac{\epsilon_{1,\perp}\kappa_{1,k}^2+\epsilon_{1,\parallel}k^2}{2\kappa_{1,k}^3}
	+
	\frac{\epsilon_{2,\perp}\kappa_{2,k}^2+\epsilon_{2,\parallel}k^2}{2\kappa_{2,k}^3}
	+
	\frac{D\beta^2k^2}{\epsilon_0(\Omega_{k}^2-\beta^2k^2)^2}.
\end{align} 
The dispersion and mode length of plasmon mode are plotted in Figure~\ref{graphene_plband}. 
We assume ZrS$_2$ as a spacer between the graphene sheet and ferromagnet and set $\epsilon_{1,\parallel}=\epsilon_{2,\parallel}=65$ and $\epsilon_{1,\perp}=\epsilon_{2,\perp}=7$~\cite{laturiaDielectricPropertiesHexagonal2018}.

\subsection{$k$-dependence of magnon-plasmon coupling}
\label{sec: k-dep coupling}
It is crucial to realize a strong magnon-plasmon coupling relative to their frequencies.
In this section, we consider the $\vec{k}$-dependence of magnon-plasmon coupling at different Fermi energies of graphene.

\begin{figure}
    \centering
    \includegraphics[width=150mm]{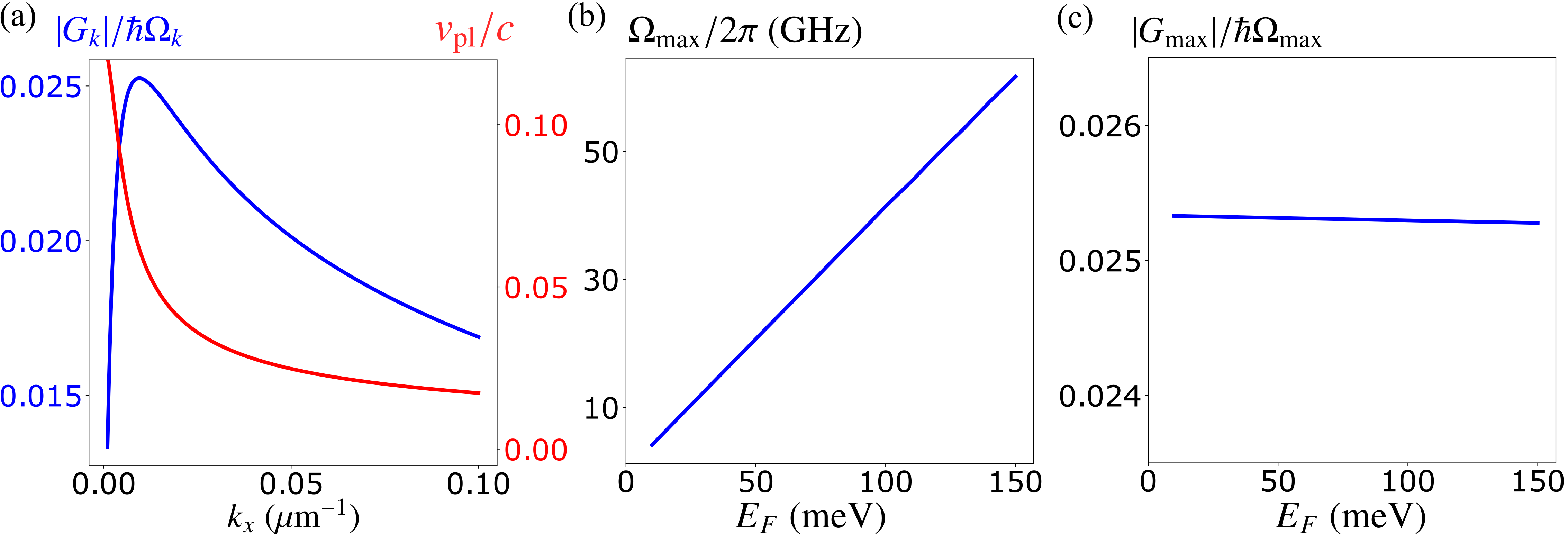}    
    \caption{
    (a) The magnon-plasmon coupling relative to their resonant frequencies and velocity of plasmon mode $v_\textrm{pl}$ are plotted as a function of wave vector $k$. Other parameters are provided in Table.~\ref{tab:parameters}. 
    (b) The frequency of plasmons at the maximum of $|G_{\vec{k}}|/\hbar\Omega_{k}$ is plotted as a function of the Fermi energy. 
    (c) The maximum value of $|G_{\vec{k}}|/\hbar\Omega_{k}$ is plotted as a function of the Fermi energy. 
     }
    \label{relative_coupling}
\end{figure}

We recall that the relative coupling strength between magnons and cavity photons is given by~\cite{soykalStrongFieldInteractions2010, Tabuchi2014, hirosawaMagnetoelectricCavityMagnonics2022}
\begin{equation}
	\frac{|G_{\vec{k}}^\textrm{cav}|}{\hbar \Omega_{k}}
	=
	\frac{g\mu_B}{2}\sqrt{\frac{\mu_0 \rho_S\eta }{\Omega_{k}}}.
\end{equation}
where $\eta$ is the volume ratio between the ferromagnetic crystal and the cavity.
We also define the spin density $\rho_S=NN_zS/V_\textrm{crys}$ with $V_\textrm{crys}$, $N$, and $N_z$ denoting the volume of magnetic crystal, the total number of spins in each layer, and the number of magnetic layers, respectively.
In the above expression, we assume the coupling with the Kittel mode.
The relative coupling strength between magnons and cavity photons depends only on $\Omega_{k}$. Thus, it is suppressed at high frequencies. 
In contrast, the relative strength of magnon-plasmon coupling is given by
\begin{equation}
	\frac{|G_{\vec{k}}|}{\hbar\Omega_{k}}
	=
	\frac{ g\mu_B\epsilon_{2,\parallel} }{2c \kappa_{2,k}}
	\sqrt{\frac{\Omega_{k}\mu_0 N_zS}{\hbar L_{k}a^2}}
	e^{-\kappa_{2,k}z},
\end{equation} 
where the coupling constant $G_{\vec{k}}$ is given in Eq.~\eqref{eq:Gk_FM}. We have assumed for simplicity that the ferromagnet and graphene have the same density~$(A=Na^2)$, i.e., for every carbon atom in graphene, there exists an atomic spin in the ferromagnet.
It is clear that the $\vec{k}$-dependence of the magnon-plasmon coupling is more complicated than the coupling with cavity photons.
Using the parameters in Table~\ref{tab:parameters}, we estimate the $\vec{k}$-dependence of the coupling strength relative to plasmon frequencies as plotted in Fig.~\ref{relative_coupling}(a).
Interestingly, $|G_{\vec{k}}|/\hbar\Omega_{k}$ strongly depends on $k_x$ with a peak at $k_x\approx0.002$~$\mu$m$^{-1}$.
Furthermore, we find that the frequency of plasmons at the maximum of $| G_{\vec{k}}|/\hbar \Omega_{k}$, denoted as $\Omega_\textrm{max}$, is proportional to $E_F$ as shown in Fig.~\ref{relative_coupling}(b). 
Therefore, it is possible to adjust $\Omega_\textrm{max}$ and enhance the magnon-plasmon coupling by tuning $E_F$.
We note that the maximum of $|G_{\vec{k}}|/\hbar\Omega_{k}$ slightly decreases with $E_F$ as shown in Fig.~\ref{relative_coupling}(c). 
Additionally, for typical parameters and a thin dielectric spacer layer with thickness $z$, we have that $\kappa_{2,k}z\ll1$. We therefore neglect in what follows the exponential decay of the magnon-plasmon coupling $\propto\exp[-\kappa_{2,k}z]$.

\section{Parameters}
In Table~\ref{tab:parameters} we list the parameters used throughout this work.
\begin{table}[h!]
    \centering
    \begin{tabular}{ccc}
        \multicolumn{3}{c}{Graphene and dielectric} \\
        \toprule
       Quantity  & Symbol & Value\\
        \hline
       Lattice constant & $a$ & \unit[2.25]{\AA} \\
       Dielectric constant (parallel) & $\epsilon_{\perp}$ & 7 \cite{laturiaDielectricPropertiesHexagonal2018}\\
       Dielectric constant (perpendicular) & $\epsilon_{\parallel}$ & 65 \cite{laturiaDielectricPropertiesHexagonal2018}\\
       Fermi velocity & $v_F$ & \unit[$10^6$]{m/s}\\
       Fermi energy & $E_F$ & \unit[100]{meV}\\
       \botrule
    \end{tabular} 
    \hfill
    \begin{tabular}{ccc}
        \multicolumn{3}{c}{Ferromagnet} \\
        \toprule
       Quantity & Symbol & Value\\
        \hline
       Lattice constant & $a$ & \unit[2.25]{\AA} \\
       Number of magnetic layers & $N_z$ & 1000 \\
       Spin quantum number & $S$ & 5/2 \\
       Ferromagnetic exchange constant & $J$ & \unit[1.15]{meV}\\
              Magnetic anisotropy & $K$ & \unit[0.04]{meV} \\
       \botrule
    \end{tabular} 

    \vspace{1cm}

    \begin{tabular}{ccc}
        \multicolumn{3}{c}{Antiferromagnet} \\
        \toprule
       Quantity & Symbol & Value\\
        \hline
       Lattice constant & $a$ & \unit[2.25]{\AA} \\
       Number of magnetic layers & $N_z$ & 1000 \\
       Spin quantum number & $S$ & 5/2 \\
       Antiferromagnetic exchange constant & $J$ & \unit[1.15]{meV}\\
      Magnetic anisotropy & $K$ & \unit[0.04]{meV} \\
       \botrule
    \end{tabular} 
    \hfill
    \begin{tabular}{ccc}
        \multicolumn{3}{c}{Skyrmion crystal} \\
        \toprule
        Quantity & Symbol & Value\\
        \hline
        Lattice constant & $a$ & \unit[2.25]{\AA} \\
        Number of magnetic layers & $N_z$ & 1000 \\
        Spin quantum number & $S$ & 5/2 \\
        Exchange constant & $J$ & \unit[10]{meV}\\
        Dzyaloshinskii-Moriya & $D$ & \unit[1]{meV}\\
        Magnetic anisotropy & $K$ & \unit[0.025]{meV} \\
       \botrule
    \end{tabular} 
    \caption{Parameters used throughout this work, for the graphene and dielectric, ferromagnet, antiferromagnet and skyrmion crystal. For simplicity and ease of comparison, we have taken the same lattice constant, number of magnetic layers and spin quantum number for all three magnetic systems. }
    \label{tab:parameters}
\end{table}

\section{Ferromagnet}
We consider a ferromagnet ($i=1$ in the main text) of finite thickness $d=N_za$, where $N_z$ is the number of layers and $a$ their spacing, on a cubic lattice, described by the Heisenberg Hamiltonian
\begin{equation}
    H_{\mathrm{magnet}}^{(1)} = -\frac{J}{2}\sum_{\langle ij \rangle} \bm S_i \cdot \bm S_j - K\sum_i (S_i^z)^2,
\end{equation}
where $J>0$ is the nearest-neighbor exchange coupling and $K>0$ is the uniaxial anisotropy. 
Introducing the truncated Holstein-Primakoff transformation as $S_i^+\approx\sqrt{2S}b_i$ and $S_i^z=S-b_i^\dagger b_i$, and applying the Fourier transforming in the plane as $b_{\bm k}={1}/{\sqrt{N}}\sum_{\bm k} e^{\textrm{i}\bm r_i \cdot \bm k} b_i$, we obtain the quadratic Hamiltonian, which describes the magnon dynamics, as
\begin{equation}
    H_{\mathrm{magnet}}^{(1)}= \hbar\sum_{\bm k}\omega_{\bm k}b_{\bm k}^\dagger b_{\bm k},
\end{equation}
where $\hbar\omega_{\bm k} = \Delta + J S(4 - 2\cos k_x a - 2\cos k_y a)$, with $\Delta\equiv2KS$ the spin-wave gap, which is $\Delta=\unit[0.2]{meV}$ for the parameters as given in Table~\ref{tab:parameters}. Here, $\bm k$ lies in the plane and we have made the uniform mode approximation, i.e., we only consider the Kittel mode. We demonstrate in Sec.~\ref{sec:Kittel} the correctness of this assumption.

We consider the coupling to the adjacent bosonic modes through the Zeeman interaction of the spins with the magnetic field generated by the plasmonic modes, 
\begin{equation}
    H_\textrm{coupling} = g \mu_B \sum_i \bm B_{p}(\bm r_i) \cdot \bm S_i , \label{eq:Hcoupling}
\end{equation}
where $g$ is the g-factor, $\mu_B$ is the Bohr magneton and $\bm B_p(\bm r_i)$ is given in Eq.~\eqref{eq:Bfield1}. In what follows, we assume that the penetration depth of the magnetic field is larger than the thickness of the film, such that we take the magnetic field $\bm B_p(\bm r_i)$ to be constant along the thickness direction.

After applying the Holstein-Primakoff transformation and Fourier transforming, the coupling between the magnon Kittel mode and the bosonic modes reads, up to linear order in magnon operators,
\begin{equation}
    H_\textrm{coupling} = \sum_{\bm k} G_{ k} \left(  \mathrm{e}^{\mathrm{i}\varphi_{\vec{k}}}
        b_{\bm k}^\dagger + \mathrm{e}^{-\mathrm{i}\varphi_{\vec{k}}} b_{-\bm k} \right) \left(a_{\bm k} + a_{\bm k}^\dagger \right)
\end{equation}
where we have disregarded constant energy shifts. Here, we have defined the coupling constant
\begin{equation}
    G_{ k} = -\mathrm{i} g \mu_B \sqrt{\frac{S N N_{z}}{2}} \mathcal B_{k}(z) ,
    \label{eq:Gk_FM}
\end{equation}
where $N_{z}$ is the number of layers in the $z$-direction and $\mathcal B_{ k}$ is defined in Eq.~\eqref{eq:Bfield2}. Note that $\mathcal B_{ k}$ is proportional to $\sqrt{1/A}=\sqrt{1/Na^2}$ and thus the factor $\sqrt{N}$ will cancel. 

We now drop the anomalous couplings, since typically the interaction strength $|G_{\bm k}| \ll\hbar\omega_c$, where $\omega_c$ is the frequency where the bands cross. We then obtain the two-boson Hamiltonian
\begin{equation}
    H^{(1)}= \sum_{\bm k} \Psi_{\bm k}^\dagger\tilde H_{\vec{k}} \Psi_{\bm k}; \quad \tilde H_{\vec{k}}= \begin{pmatrix}
        \hbar\omega_{\bm k} &  G_{ k}\mathrm{e}^{\mathrm{i}\varphi_{\vec{k}}} \\
        G_{ k} \mathrm{e}^{-\mathrm{i}\varphi_{\vec{k}}} & \hbar\Omega_{\bm k}
    \end{pmatrix} ;\quad \Psi_{\bm k} = \begin{pmatrix}
        b_{\bm k} \\ a_{\bm k}    \end{pmatrix}. \label{eq:HFM}
\end{equation}

The coupling between the modes opens up an avoided crossing, with a gap size $2|G_{k_c}|$, where $\bm k_c$ is the momentum where the uncoupled bands cross. 

\subsection{Berry curvature and Chern number}

We note that $G_{\vec{k}} = G_k$, $\omega_{\vec{k}} = \omega_{k}$ and $\Omega_{\vec{k}} = \Omega_{k}$ at long wavelengths, such that the eigenenergies only depend on the absolute value $k = |\vec{k}|$. This means that we can find a transformation such that
\begin{align}
    \tilde H_{\vec{k}}
    =
    \begin{pmatrix}
        \hbar\omega_{k} &  G_{ k}\mathrm{e}^{\mathrm{i}\varphi_{\vec{k}}} \\
        G_{ k}\mathrm{e}^{-\mathrm{i}\varphi_{\vec{k}}} & \hbar\Omega_{ k}
    \end{pmatrix}
    =
    \mathcal{U}_{\vec{k}} \mathcal{U}^{\dagger}_{\vec{k}} \begin{pmatrix}
        \hbar\omega_{ k} &  G_{ k}\mathrm{e}^{\mathrm{i}\varphi_{\vec{k}}} \\
        G_{ k}\mathrm{e}^{-\mathrm{i}\varphi_{\vec{k}}} & \hbar\Omega_{ k}
    \end{pmatrix} 
    \mathcal{U}_{\vec{k}}
    \mathcal{U}^{\dagger}_{\vec{k}}
    =
    \mathcal{U}_{\vec{k}}
    \mathcal{H}_k
    \mathcal{U}^{\dagger}_{\vec{k}},
\end{align}
with
\begin{align}
    \mathcal{H}_k = \mathcal{U}^{-1}_{\vec{k}} H_{\vec{k}} \mathcal{U}_{\vec{k}}
    &=
    \begin{pmatrix}
        \hbar \omega_{k} & G_{k}\\
        G_{k} & \hbar \Omega_{k} 
    \end{pmatrix}
    ,
    \quad
    \mathcal{U}_{\vec{k}}
    =
    \begin{pmatrix}
        \mathrm{e}^{\mathrm{i} \varphi_{\vec{k}}/2} & 0 \\
        0 & \mathrm{e}^{-\mathrm{i} \varphi_{\vec{k}}/2}
    \end{pmatrix}.
\end{align}
The eigenvectors of $\mathcal{H}_k$ and $H_{\vec{k}}$ are related as follows. If matrix $V_{k}$ diagonalizes $\mathcal{H}_{k}$, then $W_{\vec{k}} = \mathcal{U}_{\vec{k}} V_{k}$ diagonalizes $H_{\vec{k}}$. Thus, the eigenvectors are related by $\vec{w}^\pm_{\vec{k}} = \mathcal{U}_{\vec{k}} \vec{v}^\pm_k$, where $\pm$ refers to the upper and lower eigenmode.

As a result, the Berry curvature may be expressed as
\begin{align}
    F^\pm_{\vec{k}} 
    &= 
    \mathrm{i}
    \left(
        \frac{\partial\vec{w}_{\vec{k}}^{\pm,\dagger}}{\partial k_x}
        \frac{\partial\vec{w}_{\vec{k}}^\pm}{\partial k_y}
        -
        \frac{\partial \vec{w}_{\vec{k}}^{\pm,\dagger}}{\partial k_y}
        \frac{\partial\vec{w}_{\vec{k}}^\pm}{\partial k_x}
    \right)
    \nonumber \\
    &= 
    \mathrm{i}
    \left[
        \frac{\partial  \vec{v}_{k}^{\pm,\dagger} \mathcal{U}^\dagger_{\vec{k}}}{\partial k_x}
        \frac{\partial \mathcal{U}_{\vec{k}}\vec{v}_{k}^\pm}{\partial k_y}
        -
        \frac{\partial \vec{v}_{k}^{\pm,\dagger} \mathcal{U}^\dagger_{\vec{k}}}{\partial k_y}
        \frac{\partial \mathcal{U}_{\vec{k}} \vec{v}_{k}^\pm}{\partial k_x}
    \right]
    \nonumber \\
    &= 
    -\frac{1}{2k} \frac{\partial}{\partial k} \left(
        \vec{v}_{k}^{\pm,\dagger} \sigma_3   \vec{v}_{k}^{\pm} 
        \right), \label{eq:Fk-berry-Alex}
\end{align}
i.e., it also depends only on the absolute value of $\vec{k}$. We use $\sigma_3 = \text{diag}(1,-1)$ to denote the third Pauli matrix.
The Chern number thus reads
\begin{align}
    C^\pm
    &= 
    \frac{1}{2\pi} \int F^\pm_{\vec{k}} \, \mathrm{d}^2 k
    \\
    &=
    -\frac{1}{2}
    \int_0^\infty 
    \frac{\partial }{\partial k} \left( \vec{v}_{k}^{\pm,\dagger} \sigma_3 \vec{v}_{k}^{\pm} \right)
    \, \mathrm{d}k
    \\
    &=
    -\frac{1}{2}
    \left[
        \left( \vec{v}_{k}^{\pm,\dagger} \sigma_3 \vec{v}_{k}^{\pm} \right)_{k \to \infty}
        -
        \left( \vec{v}_{k}^{\pm,\dagger} \sigma_3 \vec{v}_{k}^{\pm} \right)_{k \to 0}
    \right].
    \label{eq:Chernnumber}
\end{align}

To evaluate Eq.~\eqref{eq:Chernnumber}, we first need the eigenvectors
\begin{align}
    \vec{v}_{k}^{\pm} = \frac{1}{N^\pm_k} \begin{pmatrix}
        \frac{1}{G_k} \left( \hbar \omega_k - \hbar \Omega_{k} \pm \sqrt{
           (\hbar \omega_k - \hbar \Omega_{k})^2+ 4 G_k^2
        } \right)
        \\
        1
    \end{pmatrix},
\end{align}
with the normalization constant $N^\pm_k$; we obtain
\begin{align}
    \vec{v}_{k}^{\pm,\dagger} \sigma_3 \vec{v}_{k}^{\pm}
    =
    \pm \frac{\hbar \omega_k - \hbar \Omega_{k}}{\sqrt{
           (\hbar \omega_k - \hbar \Omega_{k})^2+ 4 G_k^2
        }}.
\end{align}
For $k \to 0$, we find
\begin{align}
    \left( \vec{v}_{k}^{\pm,\dagger} \sigma_3 \vec{v}_{k}^{\pm} \right)_{k \to 0}
    =
    \pm 1,
\end{align}
and for $k \to \infty$, 
\begin{align}
    \left( \vec{v}_{k}^{\pm,\dagger} \sigma_3 \vec{v}_{k}^{\pm} \right)_{k \to \infty}
    =
    \mp 1.
\end{align}
As a result, we obtain for the Chern number
\begin{align}
    C^\pm
    =
    -\frac{1}{2} \left[ \mp 1 - (\pm 1) \right]
    =
    \pm 1.
\end{align}

\subsection{Magnon thermal Hall effect}
The finite Berry curvature leads to a finite anomalous velocity of magnon-plasmon hybrid wavepackets and, hence, to a thermal Hall effect \cite{Matsumoto2011a, Matsumoto2011b, Matsumoto2014}. The thermal Hall conductivity $\kappa_{xy}$ relates a transverse heat current density $\bm j_Q$ to a temperature gradient $\vec{\nabla} T$ according to
\begin{equation}
    \bm j_Q = -\kappa_{xy} \bm z \times \nabla T
\end{equation}
with a thermal conductivity given by \cite{Murakami2017} 
\begin{equation}
    \kappa_{xy} = -\frac{k_B^2 T}{\hbar A} \sum_{\sigma = \pm} \sum_{\bm k} \left(c_2(\rho_\sigma) - \frac{\pi^2}{3}\right) F_{\bm k}^\sigma,\label{eq:Hallkappa}
\end{equation}
where $\rho_n = f_{BE}(\epsilon_{\bm k, n})$, with $f_{BE}(x)=1/(e^{x/k_BT}-1)$, $c_2(x) = (1+x)\left(\ln\frac{1-x}{x}\right)^2 - (\ln x)^2 - 2\Li_2(-x)$ and we numerically calculate the Berry curvature as
\begin{equation}
    F_{\bm k}^\sigma \equiv \textrm{i}\sum_{\sigma \neq \sigma'} \frac{\langle \sigma |\partial_{k_x}H|\sigma'\rangle\langle \sigma'|\partial_{k_y}H|\sigma\rangle -\langle \sigma |\partial_{k_y}H|\sigma'\rangle\langle \sigma'|\partial_{k_x}H|\sigma\rangle }{( \varepsilon_{\sigma'}- \varepsilon_{\sigma})^2}. \label{eq:Berry-numerics}
\end{equation}

\subsection{Coupling to the Kittel-like mode}
\label{sec:Kittel}
As was discussed in the main text, we assume that a single plasmon mode couples to the Kittel-like mode formed by $N_z$ magnetic layers, which enhances the coupling by $\sqrt{N_z}$. Here we show that this is a correct approximation, by considering the full Hamiltonian formed by $N_z$ layers, defined as 
\begin{equation}
    H_{\mathrm{magnet}}^{(1)} = \sum_{i}\sum_{\bm k} \hbar\omega_k b_{\bm ki}^\dagger b_{\bm ki} - J S\sum_{ij} (\delta_{ij+1} +\delta_{i+1j}) b_{\bm ki}^\dagger b_{\bm kj} 
\end{equation}
where $\hbar\omega_{\bm k} = \Delta + J S(6 - \delta_{i1} - \delta_{iN} -  2\cos k_x a - 2\cos k_y a)$ and $b_{\bm ki}^\dagger$ is the creation operator in layer $i$. We assume the coupling to the magnetic field of the plasmon to be constant throughout the film, such that the coupling Hamiltonian is
\begin{equation}
    H_\textrm{coupling} = \sum_i \sum_{\bm k} G_{\bm k}' a_{\bm k}^\dagger b_{\bm k i} + \hc
\end{equation}
where we have dropped anomalous couplings and 
\begin{equation}
    G_{\bm k}' = -g \mu_B \sqrt{\frac{S N}{2}} \mathcal B_{\bm{k}}(z)  \mathrm{e}^{\mathrm{i}\varphi_{\vec{k}}}
\end{equation}
is the layer specific coupling, which importantly does not contain $\sqrt N_z$ yet---since we have not made the assumption that we only couple to the Kittel mode. 

\begin{figure}
    \centering
    \includegraphics[width=0.5\linewidth]{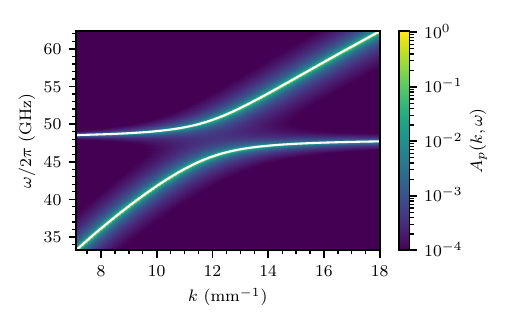}
    \caption{The plasmon spectral density defined in Eq.~\eqref{eq:plasmon-spectral-density} for $N_z=1000$ layers. The white lines are the solutions which follow from the simplified coupled mode Hamiltonian, Eq.~\eqref{eq:HFM}.}
    \label{fig:FM-layered}
\end{figure}

The total Hamiltonian is then
\begin{equation}
    H^{(1)} = \sum_{\bm k}\begin{pmatrix}
        a_{\bm k}^\dagger & \psi^\dagger_{\bm k}
    \end{pmatrix}
    \begin{pmatrix}
        \hbar\Omega_{\bm k} & \bar G_{\bm k}' \\
        ({\bar G_{\bm k}'})^\dagger &\bar H_{\bm k}
    \end{pmatrix}
    \begin{pmatrix}
        a_{\bm k} \\ \psi_{\bm k}
    \end{pmatrix} \label{eq:HFM-layered}.
\end{equation}
Here, $\psi_{\bm k}^\dagger=(b_{\bm k1}^\dagger\dots b_{\bm k N_z}^\dagger)^T$ is a vector containing $N_z$ magnon operators, $[\bar G_{\bm k}']_{ij}=G_k'$ is the coupling Hamiltonian matrix and $[\bar H_{\bm k}]_{ij}=\hbar\omega_{\bm k} - J  S( \delta_{ij+1} +\delta_{i+1j})$ is the magnon Hamiltonian matrix. We now diagonalize the Hamiltonian in Eq.~\eqref{eq:HFM-layered} and show in Fig.~\ref{fig:FM-layered} the plasmon spectral density, defined as 
\begin{equation}
    A_p(\bm k,\omega) = \sum_i |\psi_{pi}|^2\delta(\omega-\omega_{\bm ki}), \label{eq:plasmon-spectral-density}
\end{equation}
where $\psi_{pi}$ is the plasmon weight of the $i$-th eigenvector and $\omega_{\bm k i}$ its corresponding eigenfrequency.\footnote{In the ordering of  Eq.~\eqref{eq:HFM-layered}, the plasmon weight is the first item of the eigenvector.} The white lines are the frequencies which follow from diagonalizing the simplified coupled-mode Hamiltonian, Eq.~\eqref{eq:HFM}, which are also shown in Fig.~2 in the main text. Importantly, the coupling $G_{\bm k}$ that enters the simplified coupled-mode Hamiltonian is proportional to $\sqrt N_z$, since there we have made the assumption that we only couple to the Kittel-like mode.

From Fig.~\ref{fig:FM-layered} we observe that the full $(N_z+1)\times (N_z+1)$ Hamiltonian and the simplified coupled-mode Hamiltonian are in direct agreement. We therefore conclude that it is well justified to approximate the $N_z$ magnon modes as a single Kittel-like mode, and thus the coupling scales as $\sqrt N_z$.

\section{Antiferromagnet}
We now consider an antiferromagnet ($i=2$ in the main text) of thickness $d = N_z a$, where $N_z$ is the number of layers and $a$ their spacing, on a square lattice, described by the Heisenberg Hamiltonian,
\begin{equation}
    H_{\mathrm{magnet}}^{(2)} = \frac{J}{2}\sum_{\langle ij\rangle} \bm S_i \cdot \bm S_j - K\sum_i (S_i^z)^2,
\end{equation}
where $J>0$ is the nearest-neighbor antiferromagnetic exchange coupling and $K>0$ is the uniaxial anistropy constant. In absence of a magnetic field, the magnetic ground state is an antiferromagnetic ordering along the $z$ axis.

We now introduce the Holstein-Primakoff operators for the $A/B$ sublattice as 
\begin{equation}
    S_{A;i}^+\approx\sqrt{2S}b_{1,i};\quad S_{B;i}^+\approx\sqrt{2S}b_{2,i}^\dagger;\quad S_{A;i}^z=S-b_{1,i}^\dagger b_{1,i};\quad  S_{B;i}^z=-S+b_{2,i}^\dagger b_{2,i}, \label{eq:HP-AFM}
\end{equation}
where $b_{1,i}^\dagger, b_{1,i}$ and $b_{2,i}^\dagger,b_{2,i}$ are the creation and annihilation operators for spin deviations at sites $A$ and $B$ in unit cell $i$ respectively. After the Fourier transformation
\begin{equation}
    b_{1,i} = \frac{1}{\sqrt{N}} \sum_{\bm k} e^{\textrm{i}\bm k\cdot \bm r_i} b_{1,\bm k};\quad b_{2,i} = \frac{1}{\sqrt{N}} \sum_{\bm k} e^{\textrm{i}\bm k\cdot \bm r_i} b_{2,\bm k}
\end{equation}
we obtain the quadratic part of the spin-wave Hamiltonian as

\begin{equation}
    H_{\mathrm{magnet}}^{(2)} =  \sum_{\bm k} A_{\bm k} (b_{1,\bm k}^\dagger b_{1,\bm k} + b_{2,\bm k}^\dagger b_{2,\bm k}) + \frac12   B_{\bm k} (b_{1,\bm k} b_{2,-\bm k} + b_{1,\bm k}^\dagger b_{2,-\bm k}^\dagger) 
    \label{eq:ham:afm}
\end{equation}
where $A_{\bm k} = JSz + 2SK$ and $B_{\bm k}=\gamma_{\bm k} JSz$. 
Here $z=4$ is the coordination number, i.e., the number of nearest neighbours and $\gamma_{\bm k}\equiv (1/z)\sum_\delta e^{i\bm k \cdot \bm\delta}$ is the structure factor, with $\bm \delta$ the vectors connecting the nearest neighbors. For a simple square lattice with the distance between nearest neighbors $a$ we consider here, $\gamma_{\bm k} = \frac12\cos(k_x a)+\frac12\cos(k_y a)$.

To diagonalize the Hamiltonian, Eq.~\eqref{eq:ham:afm}, we perform a Bogoliubov transformation with 
\begin{align}
    b_{1,\bm k} &= \cosh\phi\, \alpha_{\bm k} + \sinh\phi\, \beta_{-\bm k}^\dagger \\
    b_{2,-\bm k}^\dagger &= \sinh\phi\, \alpha_{\bm k} + \cosh\phi\, \beta_{-\bm k}^\dagger
\end{align}
where $\tanh 2 \phi = -\frac{B_{\bm k}}{A_{\bm k}}$,
such that 
\begin{equation}
    H_{\mathrm{magnet}}^{(2)} = \hbar \sum_{\bm k} \omega_{\bm k} (\alpha_{\bm k}^\dagger \alpha_{\bm k} + \beta_{\bm k}^\dagger \beta_{\bm k});\quad \hbar\omega_{\bm k} = \sqrt{A_{\bm k}^2-B_{\bm k}^2}.
\end{equation}
Furthermore, the magnon gap is given by $\hbar\omega_0\equiv S\sqrt{2SK(2SK+JSz)}$, which for the parameters used in this work is $\hbar\omega_0=\unit[2.15]{meV}$.

The coupling to the plasmon mode is again given by Eq.~\eqref{eq:Hcoupling}, which becomes after introduction of the Holstein-Primakoff operators given in Eq.~\eqref{eq:HP-AFM} and applying the Boguliobov transformation, 
\begin{align}
    H_\textrm{coupling} &=  \sum_{\bm k } a_{\bm k } \left[ G_{k} \mathrm{e}^{\mathrm{i}\varphi_{\vec{k}}}(b_{1,-\bm k}+b_{2,\bm k}^\dagger)+ G_{ k } \mathrm{e}^{-\mathrm{i}\varphi_{\vec{k}}}(b_{2,-\bm k}+b_{1,\bm k}^\dagger)  \right] + \hc \\
    &= \sum_{\bm k }a_{\bm k } \left[ \tilde G_{\bm k } (\alpha_{-\bm k } +\beta^\dagger_{\bm k })+\tilde G_{\bm k }^*(\alpha_{\bm k }^\dagger +\beta_{-\bm k }) \right] + \hc 
\end{align}
where $\tilde G_{\bm k} = f_{\bm k}G_{ k}\mathrm{e}^{\mathrm{i}\varphi_{\vec{k}}}$, with $f_{\bm k}\equiv \cosh\phi+\sinh\phi$ an exchange enhancement factor \cite{Dyrdal2023}.

Assuming that the interaction strength is smaller than the band crossing energy, the anomalous couplings can again be dropped and we can write this as the effective three-boson Hamiltonian
\begin{equation}
    H^{(2)} = \sum_{\bm k} \Psi_{\bm k}^\dagger \tilde H_{\bm k} \Psi_{\bm k}; \quad \Psi_{\bm k} = \begin{pmatrix}
        \alpha_{\bm k} \\ \beta_{\bm k} \\ a_{\bm k}    \end{pmatrix}, 
        \label{eq:Ham-AFM}
\end{equation}
with Hamiltonian matrix
\begin{align}
    \tilde H_{\bm k}
    =
    \begin{pmatrix}
        \hbar\omega_{\bm k} & 0 & \tilde G_{\bm k} \\
        0 & \hbar\omega_{\bm k} & \tilde G_{\bm k}^* \\
        \tilde G_{\bm k}^* & \tilde G_{\bm k} & \hbar\Omega_{\bm k}
    \end{pmatrix}.
\end{align}
Due to the arrowhead structure of the Hamilton matrix one eigenvalue is the magnon energy itself, $\hbar\omega_{\bm k}$, and the other two eigenvalues correspond to anticrossing modes.

Using $\hbar\omega_{\vec k} = \hbar\omega_{-\bm k}$, $\hbar\Omega_{\bm k} = \hbar\Omega_{-\bm k}$, and $G_{\bm k}^* = G_{-\bm k}$, one can show that $\tilde H^\ast_{\bm k} = \tilde H_{-\bm k}$ and $\tilde H_{\bm k} = U^\dagger \tilde H_{-\bm k} U$, with
\begin{align}
    U = \begin{pmatrix}
        0 & 1 & 0 \\
        1 & 0 & 0 \\
        0 & 0 & 1
    \end{pmatrix}.
\end{align}
Thus, the system holds both time-reversal and space inversion symmetry, implying that the Berry curvature vanishes identically.

\subsection{Magnon spin Nernst effect}
As the Berry curvature vanishes, the thermal Hall response is zero. However, the opposite chirality of the branches allows for a magnon spin Nernst effect, where an applied temperature gradient induces a transverse magnon spin current as \cite{Cheng2016SNE,Zyuzin2016SNE} 
\begin{equation}
    \bm j_m = -\frac{k_B}{A\hbar} \bm z \times\nabla T \sum_n\sum_{\bm k} \Omega_{S_z}^n c_1(\rho_n);\quad c_1(\rho_n)\equiv  \rho_n\ln\rho_n-(1+\rho_n)\ln(1+\rho_n), \label{
    eq:jm-nernst}
\end{equation}
where we defined the magnon spin Berry curvature \cite{To2023}
\begin{equation}
    \Omega_{S_z}^n \equiv i\sum_m \frac{\langle n |J_{x}^{S^z}|m\rangle\langle m|\partial_{k_y}H|n\rangle -\langle n |J_{y}^{S^z}|m\rangle\langle m|\partial_{k_x}H|n\rangle }{( E_m- E_n)^2}, \label{eq:nernstBerry}
\end{equation}
and $J^{S^z}_{\alpha}= \hbar^{-1}S^z \partial_{k_\alpha} H + \hbar^{-1}\partial_{k_\alpha} H  S^z $ is the spin current tensor operator with $S^z=\hbar\Diag[1,-1,0]$ the $S^z$ operator. Additionally, $|m\rangle$ and $E_m$ are the eigenvectors and eigenvalues obtained form diagonalizing the Hamiltonian.
We can now identify the magnon Nernst coefficient from  Eq.~\eqref{eq:jm-nernst} as  
\begin{equation}
    \alpha_{xy} = \frac{k_B}{A\hbar} \sum_n\sum_{\bm k} c_1(\rho_n)\Omega_{S_z}^n \label{eq:Nernsteta}
\end{equation}
such that $\vec{j}_m=-\alpha_{xy}\vec{z}\times\nabla T$. 

To calculate the magnon spin Nernst coefficient requires the numerical evaluation of the magnon spin Berry curvature, Eq.~\eqref{eq:nernstBerry}. To do this, we perform the derivatives through a finite difference scheme, 
\begin{equation}
    \frac{\partial H}{\partial k_x} = \frac{H|_{k_x+\Delta k} - H|_{k_x-\Delta k}}{2\Delta k};\quad \frac{\partial H}{\partial k_y} = \frac{H|_{k_y+\Delta k} - H|_{k_y-\Delta k}}{2\Delta k}.
\end{equation}
We choose $\Delta k = \unit[0.01]{\mu m^{-1}}$, which leads to sufficiently converged results.
For numerical stability, we have to include a tiny splitting of the magnons at $k=0$, which can be understood as applying a small magnetic field of $H=\unit[10]{\mu T}$.

\section{Anomalous couplings in the thermal Hall and magnon spin Nernst effects}
In the main text, we have dropped the anomalous coefficients in the calculations of the thermal Hall and magnon spin Nernst coefficients. This was motivated by the fact that $|G_{\bm{k}_c}|\ll\hbar\omega_c$, i.e., the anomalous couplings at the band crossing are much smaller than the energy at which the crossing occurs. The anomalous couplings thus have negligible effect on the spectral properties of the hybridization of the magnon and plasmon quasiparticles. It is not \textit{a priori} clear that the same holds for the transport coefficients, and we thus restore in this section the anomalous couplings, and recalculate the thermal Hall and magnon spin Nernst coefficients. 

The thermal Hall conductivity is still given by Eq.~\eqref{eq:Hallkappa}, but now the Berry curvature is given by \cite{Shindou13,Matsumoto2014}
\begin{equation}
    F_{\bm k,n} \equiv \textrm{i}\sum_{m \neq n} \nu_m\nu_n \frac{\langle n |\partial_{k_x}H|m\rangle\langle m|\partial_{k_y}H|n\rangle -\langle n|\partial_{k_y}H|m\rangle\langle m|\partial_{k_x}H|n\rangle }{(\nu_m E_m- \nu_n E_n)^2}. \label{eq:hallBerryFull}
\end{equation}
Similarily, the Nernst conductivity is given by Eq.~\eqref{eq:Nernsteta}, but with
\begin{equation}
    \Omega_{S_z}^n \equiv \textrm{i}\sum_{m\neq n} \nu_m\nu_n \frac{\langle n |J_{x}^{S^z}|m\rangle\langle m|\partial_{k_y}H|m\rangle -\langle n |J_{y}^{S^z}|m\rangle\langle m|\partial_{k_x}H|n\rangle }{(\nu_m E_m- \nu_n E_n)^2}. \label{eq:nernstBerryFull}
\end{equation}
Here $J^{S^z}_{\alpha}= S^z \nu\partial_{k_\alpha} H + \partial_{k_\alpha} H \nu S^z $ is the spin current tensor operator, with $\nu = \Diag[1,1,1,-1,-1,-1]$ and $S^z=\Diag[1,-1,0,1,-1,0]$ the $S^z$ operator \cite{To2023}. 
Additionally, $|n\rangle$ and $E_n$ are the eigenstates obtained from diagonalizing the full Hamiltonian using Colpa's diagonalization algorithm \cite{Colpa1978}. Importantly, the sums over $n$ in Eqs.~(\ref{eq:Hallkappa}, \ref{eq:Nernsteta}) still only run over the first $N$ bands (only the ``particle'' bands), and not over the full $2N$ bands.

\begin{figure}
    \centering
    \begin{tikzpicture}
            \node (A) at (-0.5\textwidth,0) {\includegraphics[width=0.5\textwidth]{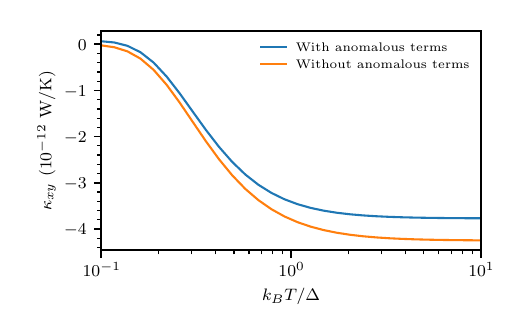}};
			\node (B) at (0.0\columnwidth,0) {\includegraphics[width=0.5\textwidth]{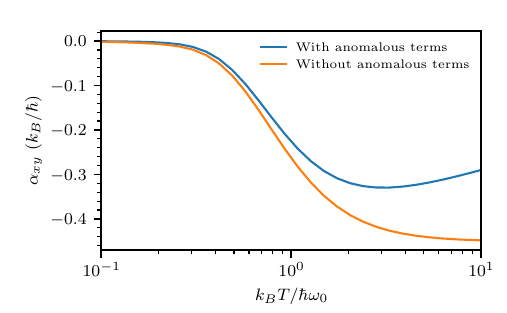}};
			\node [align=center] at ([shift={(A)}] 0.5,2.5) {(a) Ferromagnet};
			\node [align=center] at ([shift={(B)}] 0.5,2.5) {(b) Antiferromagnet};
    \end{tikzpicture}
    \caption{Comparing the effect of the anomalous couplings on the: (a) thermal Hall coefficient as a function of temperature for the ferromagnet-graphene structure; (b) magnon Nernst coefficient as a function of temperature for the antiferromagnet-graphene structure.}
    \label{fig:anomalous}
\end{figure}

We now calculate the thermal Hall and magnon spin Nernst conductivities, comparing the cases with and without anomalous coefficients in Fig.~\ref{fig:anomalous}. Firstly, we conclude that the thermal Hall conductivity, $\kappa_{xy}$, in the ferromagnetic case shows little effect of the inclusion of the anomalous coefficients. 
The magnon spin Nernst conductivity in the antiferromagnetic case is however noticably affected by the inclusion of the anomalous coefficients at temperatures $k_\text{B}T/\hbar \omega_0$. In particular, instead of saturating for temperatures $k_BT>\hbar\omega_0$, the spin Nernst conductivity decreases.

\section{Skyrmion crystals}

We consider the following spin-lattice Hamiltonian on a square lattice:
\begin{equation}
	H^{(3)}_\textrm{magnet}
	=
	-\frac{J}{2}\sum_{\braket{i,j}}\vec{S}_{i}\cdot \vec{S}_{j}+\frac{1}{2}\sum_{\braket{i,j}}\vec{D}_{i,j}\cdot\vec{S}_{i}\times\vec{S}_{j}
	+g\mu_{B} B_z \sum_{i} \vec{S}_{i}\cdot\hat{\vec{z}} -K\sum_{i} (S_{i}^z)^2\,,
\end{equation}
with the ferromagnetic exchange ($J>0$), interfacial Dzyaloshinskii-Moriya interaction ($\vec{D}_{i,j}=D\hat{z}\times (\vec{r}_i-\vec{r}_j)/|\vec{r}_i-\vec{r}_j|$), and easy-axis anisotropy ($K>0$).
The above Hamiltonian stabilizes skyrmion crystals~(SkXs) at finite magnetic fields for $KJ/D^2<1.8$~\cite{wilsonChiralSkyrmionsCubic2014}.
We set $D/J=0.5$ and $KJ/D^2=0.25$ in the following.

Since external magnetic fields would open a large cyclotron gap in graphene plasmon bands~\cite{castronetoElectronicPropertiesGraphene2009}, we study the magnon spectrum of the metastable zero-field SkX using linear spin-wave theory. The classical configuration of magnetic moments is obtained by the Monte Carlo annealing and Landau–Lifshitz–Gilbert~(LLG) simulation, as shown in Fig.~\ref{Fit_dispersion}(a). 
We note that the direction of magnetic moments is opposite to spins due to the negative charge of electrons. 
Denoting the normalized classical spins as $\hat{\vec{e}}_{i}^z$, we introduce the local orthogonal basis $(\hat{\vec{e}}_{i}^x,\hat{\vec{e}}_{i}^y,\hat{\vec{e}}_{i}^z)$. 
To quantize spin excitations, we perform the Holstein-Primakoff transformation:
\begin{equation}
\vec{S}_{i}\approx 
\hat{\vec{e}}_{i}^z\left(S-b_{i}^\dagger b_{i}\right)
+\sqrt{\frac{S}{2}}\left(\hat{\vec{e}}_{i}^-b_{i}+\hat{\vec{e}}_{i}^+b_{i}^\dagger\right),
\end{equation}
where $S$ is the spin quantum number and $\hat{\vec{e}}_{i}^{\pm}=\hat{\vec{e}}_{i}^x\pm \textrm{i}\hat{\vec{e}}_{i}^y$.
The Fourier transform is performed as $b^\dagger_{i,j}=\frac{1}{\sqrt{N_\textrm{u}}}\sum_{\vec{k}}e^{-\textrm{i}\vec{k}\cdot\vec{r}_{i,j}}b^\dagger_{\vec{k},j}$, where $i$ and $j$ label the unit cell and basis atom with $\vec{r}_{i,j}=\vec{R}_i+\vec{r}_j$, respectively. Here, $N_\textrm{u}$ denotes the number of unit cells. The total number of spins is given as $N=N_\textrm{u}\times N_\textrm{b}$ with $N_b$ representing the number of spins within a magnetic unit cell.
The Bogoliubov-de Gennes form of spin-wave Hamiltonian is written as~\cite{Diaz2020FM,Hirosawa2020,hirosawaLasercontrolledRealReciprocalspace2021}
\begin{equation}
    H^{(3)}_\textrm{magnet}
    =
    \frac{S}{2}\sum_{\vec{k},j,j'}\psi_{\vec{k},j}^\dagger 
    \begin{pmatrix}
    \Gamma_{jj'}(\vec{k})& \Delta_{jj'}(\vec{k})\\
    \Delta_{jj'}^*(-\vec{k})& \Gamma_{jj'}^*(-\vec{k})
    \end{pmatrix}
    \psi_{\vec{k},j'},
    \label{eq:SM_spinwave}
\end{equation}
where $\psi_{\vec{k},j}=(b_{\vec{k},j},b^\dagger_{-\vec{k},j})^\text{T}$.
We define
\begin{subequations}
\begin{align}
    \Lambda_j(\vec{k})&
    =
    \sum_{j'} [J_{jj'}(\vec{k}=0)\hat{\vec{e}}^z_j\cdot\hat{\vec{e}}_{j'}^z
    -\vec{D}_{jj'}(\vec{k}=0)\cdot\hat{\vec{e}}^z_j\times \hat{\vec{e}}_{j'}^z] 
    - K\Big\{(\hat{\vec{z}}\cdot\hat{\vec{e}}_j^x)^2+(\hat{\vec{z}}\cdot\hat{\vec{e}}_j^y)^2-2(\hat{\vec{z}}\cdot\hat{\vec{e}}_j^z)^2\Big\}-g\mu_BB_z\hat{\vec{z}}\cdot\hat{\vec{e}}_j^z,
    \\
    \Lambda'_j(\vec{k})&
    =-K\Big\{(\hat{\vec{z}}\cdot\hat{\vec{e}}_j^x)^2+2i(\hat{\vec{z}}\cdot\hat{\vec{e}}_j^x)(\hat{\vec{z}}\cdot\hat{\vec{e}}_j^y)-(\hat{\vec{z}}\cdot\hat{\vec{e}}_j^y)^2 \Big\},
    \\
    \Gamma_{jj'}(\vec{k}) &=
    \delta_{jj'}\Lambda_j+\frac{1}{2}[-J_{jj'}(\vec{k})\hat{\vec{e}}_j^+\cdot \hat{\vec{e}}_{j'}^- +\vec{D}_{jj'}(\vec{k})\cdot \hat{\vec{e}}_j^+\times \hat{\vec{e}}_{j'}^-],\\
    \Delta_{jj'}(\vec{k})&
    =
    \delta_{jj'}\Lambda'_j+\frac{1}{2}[-J_{jj'}(\vec{k})\hat{\vec{e}}_j^+\cdot \hat{\vec{e}}_{j'}^+ +\vec{D}_{jj'}(\vec{k})\cdot \hat{\vec{e}}_j^+\times \hat{\vec{e}}_{j'}^+],
\end{align}
\end{subequations}
with $J_{jj'}(\vec{k})=\sum_i J_{\vec{r}_{i,j},\vec{r}_{0,j'}} \mathrm{e}^{-\mathrm{i}\vec{k}\cdot(\vec{r}_{i,j}-\vec{r}_{0,j'})}$ and $\vec{D}_{jj'}(\vec{k})=\sum_i \vec{D}_{\vec{r}_{i,j},\vec{r}_{0,j'}} \mathrm{e}^{-\mathrm{i}\vec{k}\cdot(\vec{r}_{i,j}-\vec{r}_{0,j'})}$. The spin-wave Hamiltonian is diagonalized by the paraunitary matrix $T_{\vec{k}}$~\cite{Colpa1978}, satisfying $T^\dagger_{\vec{k}}\Sigma T_{\vec{k}}=T_{\vec{k}}\Sigma T^\dagger_{\vec{k}}=\Sigma$ where
\begin{equation}
    \Sigma=
    \begin{pmatrix}
        I_{N_b\times N_b}&0
        \\
        0 & -I_{N_b\times N_b}
    \end{pmatrix}.
\end{equation}

\subsection{Rescaled energy and magnetic field}
\label{sec:Energyscale_SkX}

We take the lattice constant and ferromagnetic exchange strength unity, so the spin-lattice model is dimensionless.
For comparison with experiments, the energy is scaled as~\cite{Diaz2020FM, hirosawaMagnetoelectricCavityMagnonics2022}
\begin{equation}
    \tilde{\mathsf{E}}_n=\frac{\tilde{J}}{\sqrt{3}J}\left(\frac{\tilde{D}}{\tilde{J}}\right)^2\left(\frac{D}{J}\right)^{-2}\mathsf{E}_n
    \label{eq:SM_tildeEn}
\end{equation}
where $D/J=0.5$ in our spin-lattice model and $\tilde{J},\tilde{D}$ are the material parameters.
In the main manuscript, we assume $\tilde{J}=10$~meV, $\tilde{D}=1$~meV, and $\tilde{K}=0.025$~meV for numerical estimates. 
With these parameters, the eigenfrequency of the CCW mode is found to be 39~GHz.
In addition, we assume the distance between skyrmions to be 100~nm.

\subsection{Topological magnons in SkXs}
\label{sec:top_mag_SkX}
The topological spin textures induce emergent magnetic fields on magnons, which leads to the nontrivial magnon band topology in SkXs~\cite{RoldanMolina2016, Diaz2020FM, Hirosawa2020, hirosawaLasercontrolledRealReciprocalspace2021}.
We denote the magnon wave function as $\ket{u_{\vec{k}}^n}=T_{\vec{k}}\vec{v}_n$ with $v_n^j=\delta_{j,n}$, where the orthogonality relation holds as $\braket{u_{\vec{k}}^m |u_{\vec{k}}^n}_\textrm{para}=\xi \vec{v}_m^T T^\dagger_{\vec{k}}\Sigma T_{\vec{k}}\vec{v}_n =\delta_{mn}$ with $\xi=\pm1$ for particle/hole bands.
The Berry curvature of magnons in noncollinear magnets is then defined as~\cite{Shindou13}
\begin{align}
   F_{\vec{k}}^n=\textrm{i}\left[\Braket{\frac{\partial u_{\vec{k}}^n }{\partial k_x} |\frac{\partial u_{\vec{k}}^n}{\partial k_y} }_\textrm{para}-\Braket{\frac{\partial u_{\vec{k}}^n }{\partial k_y} |\frac{\partial u_{\vec{k}}^n}{\partial k_x} }_\textrm{para}\right].
\end{align}
We find that the Chern number of the CCW mode is $-1$ at $B_z=0$~\cite{Fukui2005}.
From the bulk-boundary correspondence, a chiral magnonic edge state appears across the band gap above the CCW mode.

\begin{figure}
    \centering
    \includegraphics[width=150mm]{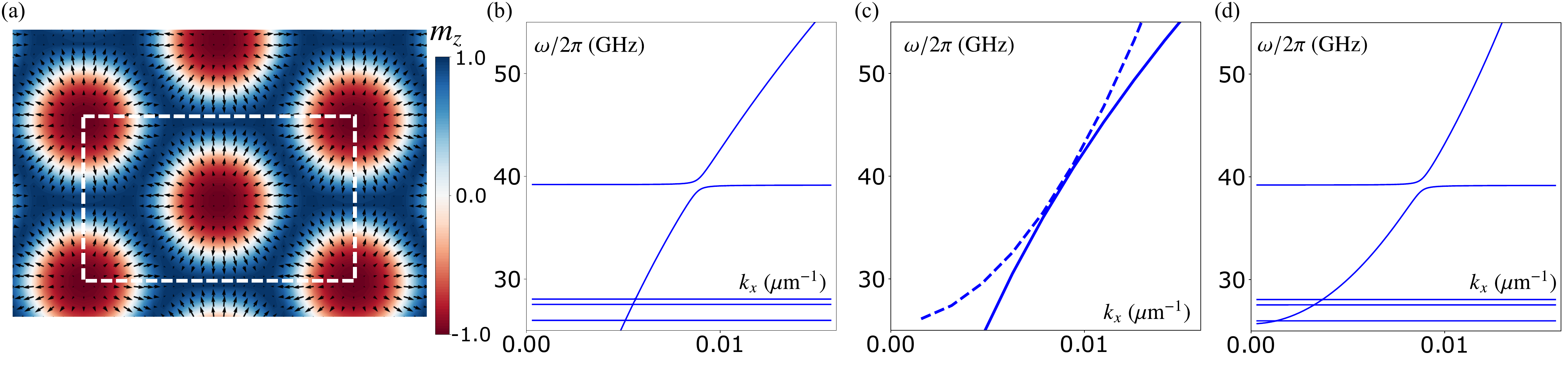}    
    \caption{
    (a)~The magnetization of skyrmion crystals is plotted ($\vec{m}_i=-\hat{\vec{e}}_{i}^z$). The magnetic unit cell consists of $27 \times 16$ spins, indicated by dashed lines.
    (b)~The magnon-plasmon hybrid spectrum is plotted near the $\Gamma$ point.
    (c)~The dispersion of plasmons is fitted by a quadratic curve. A blue solid~(dashed) line indicates the plasmon dispersion (fitted line).
    (d)~The magnon-plasmon hybrid spectrum is plotted with the fitted plasmon band. We fix $k_y=0$ in (b-d).
     }
    \label{Fit_dispersion}
\end{figure}

\subsection{Magnon-plasmon hybrids in SkXs}

We consider the magnetic dipole coupling between magnetic fields and magnets,
\begin{equation}
H_\textrm{coupling}=g\mu_B\sum_{\vec{r}} \vec{B}(\vec{r},t)\cdot \vec{S}_{\vec{r}}.
\end{equation}
Performing the Holstein-Primakoff transformation, the bilinear magnon-plasmon coupling is then written as
\begin{equation}
H_\textrm{coupling}=g\mu_B\sqrt{\frac{S}{2}}\sum_{\vec{r}} 
\vec{B}(\vec{r},t)
\cdot
\left(\hat{\vec{e}}_{\vec{r}}^-b_{\vec{r}}+\hat{\vec{e}}_{\vec{r}}^+b_{\vec{r}}^\dagger\right)
=-g\mu_B\sqrt{\frac{SN_b}{2}}\sum_{\vec{r}}
\vec{B}(\vec{r},t)
\cdot
\left(\vec{\mu}_{\vec{r}} b_{\vec{r}}+\vec{\mu}_{\vec{r}}^*b_{\vec{r}}^\dagger\right),
\end{equation}
where $\vec{\mu}_{\vec{r}}=-\hat{\vec{e}}_{\vec{r}}^-/\sqrt{N_b}$ and $\vec{\mu}_{\vec{r}}^*=-\hat{\vec{e}}_{\vec{r}}^+/\sqrt{N_b}$.
After the Fourier transform, we obtain
\begin{align}
H_\textrm{coupling}&=
-
g\mu_B\sqrt{\frac{NS}{2}}
\sum_{\vec{k}}
\sum_{j=1}^{N_\textrm{b}}
\left[
\left(\vec{B}_{\vec{k}}\cdot\vec{\mu}_j\right)a_{\vec{k}}b_{-\vec{k},j}
+\left(\vec{B}_{\vec{k}}\cdot\vec{\mu}_j^*\right)a_{\vec{k}}b_{\vec{k},j}^\dagger
+\left(\vec{B}_{\vec{k}}^*\cdot\vec{\mu}_j\right)a_{\vec{k}}^\dagger b_{\vec{k},j}
+\left(\vec{B}^*_{\vec{k}}\cdot\vec{\mu}_j^*\right)a_{\vec{k}}^\dagger b_{-\vec{k},j}^\dagger
\right]
\nonumber\\
&=
\sum_{\vec{k}}
\sum_{j=1}^{N_\textrm{b}}
\textrm{i}G_k(\vec{\hat{k}}\times \hat{z})\cdot
\left[
\vec{\mu}_j \,a_{\vec{k}}b_{-\vec{k},j}
+\vec{\mu}_j^* \, a_{\vec{k},\beta}b_{\vec{k},j}^\dagger
- \vec{\mu}_j \, a_{\vec{k}}^\dagger b_{\vec{k},j}
- \vec{\mu}_j^* \, a_{\vec{k}}^\dagger b_{-\vec{k},j}^\dagger
\right]
,\label{m-p_coupling1}
\end{align}
where $\vec{B}_{\vec{k}}=\mathcal{\vec{B}}_{k}(\vec{\hat{k}}\times \hat{z})$, 
$
\mathcal{\vec{B}}_{k}
=
-\frac{\textrm{i}\epsilon_{2,\parallel} \Omega_{k}}{c\kappa_{2,k}}\sqrt{\frac{\hbar\Omega_{k}\mu_0}{2AL_{k}}}
$, and $G_k=\frac{ g\mu_B\epsilon_{2,\parallel} \Omega_k}{2c \kappa_{2,k}}\sqrt{\frac{\hbar\Omega_{k}\mu_0 N_zS}{L_{k}a^2}}$ as given in Sec.~\ref{sec: plasmon}.
Combining Eqs.~\eqref{m-p_coupling1} and \eqref{eq:SM_spinwave}, the magnon-plasmon BdG Hamiltonian is written as
\begin{equation}
	H^{(3)}
	=
	\frac{1}{2}\sum_{\vec{k},j,j'}
	\begin{pmatrix}
	 	a^\dagger_{\vec{k}}
	 	&
	 	b_{\vec{k},j}^\dagger
	 	&
	 	a_{-\vec{k}}
	 	&
	 	b_{-\vec{k},j}
	\end{pmatrix}
	\begin{pmatrix}
		\hbar\Omega_k   
		& -\textrm{i}G_k (\vec{\hat{k}}\times \hat{z})\cdot\vec{\mu}_{j'} 
		& 0 & -\textrm{i}G_k (\vec{\hat{k}}\times \hat{z})\cdot\vec{\mu}_{j'}^*
		\\
		 \textrm{i}G_k (\vec{\hat{k}}\times \hat{z})\cdot\vec{\mu}_j^* 
		 & S\Gamma_{jj'}(\vec{k}) & \textrm{i}G_k (\vec{\hat{k}}\times \hat{z})   \cdot\vec{\mu}_{j}^*
		 &S\Delta_{jj'}(\vec{k})
		 \\
		 0 &-\textrm{i}G_k (\vec{\hat{k}}\times \hat{z})\vec{\mu}_{j'} 
		 & \hbar\Omega_k   &-\textrm{i}G_k (\vec{\hat{k}}\times \hat{z})\cdot\vec{\mu}_{j'}^*
		 \\
		 \textrm{i}G_k (\vec{\hat{k}}\times \hat{z})\cdot\vec{\mu}_{j}
		 &S\Delta_{jj'}^*(-\vec{k})
		 & \textrm{i}G_k (\vec{\hat{k}}\times \hat{z})\cdot\vec{\mu}_j 
		  & S\Gamma_{jj'}^*(-\vec{k})
	\end{pmatrix}
	\begin{pmatrix}
	 	a_{\vec{k}}
	 	\\
	 	b_{\vec{k},j'}
	 	\\
	 	a_{-\vec{k}}^\dagger
	 	\\
	 	b_{-\vec{k},j'}^\dagger
	\end{pmatrix}.
\end{equation}

As discussed in Sec.~\ref{sec: k-dep coupling}, we optimize the strength of magnon-plasmon coupling relative to their resonance frequencies by tuning the Fermi energy of graphene.
Among various magnon excitations in SkXs, the counterclockwise (CCW) and clockwise (CW) modes couple with spatially uniform in-plane magnetic fields~\cite{mochizukiSpinWaveModesTheir2012}.
Since the CCW mode has a negative mass, we set $E_F= 100$~meV to maximize the coupling strength between the CCW mode and plasmons at $\omega_{\vec{k}}\approx 40$~GHz.
The other parameters are provided in Table~\ref{tab:parameters}.
Figure~\ref{Fit_dispersion}(b) shows the magnon-plasmon hybrid spectrum in skyrmion crystals. Since the velocity of plasmons is much larger than magnons, the magnon bands show no dispersion within the plotted energy and momentum range. 
The band crossing between the CCW and plasmon modes occurs at $k_c=8.8\times 10^{-3}$~$\mu$m$^{-1}$, where the hybridization gap is about $1.1$~GHz.
At the crossing point $k=k_c$, we estimate the velocity of plasmon and CCW mode as $v_\textrm{pl}=0.066c$ and $v_\textrm{mag}=-3.3$~ms$^{-1}$, respectively. Plasmons are approximately $6\times10^6$ times faster than magnons at the band crossing. Also, the coupling strength is estimated as $G_k/h=0.98$~GHz at $k=k_c$.

To investigate the topological nature of magnon-plasmon hybrids, we construct the lattice-discretized model of plasmons.
Since the velocity of the CCW mode is several orders of magnitude slower than plasmons, it is sufficient to fit the gradient, frequency, and coupling constant only at the crossing point.
Fitting the plasmon dispersion with quadratic terms, the plasmon dispersion is approximated as
\begin{align}
    \hbar\Omega_{\vec{k}}'
    &=
    \frac{k_x^2  +k_y^2}{2M_\textrm{pl}^2 a^2}+E_\textrm{pl}
    \approx 
    \frac{2(1-\cos k_x N_x)}{2M_\textrm{pl}^2 L_x^2}+
    \frac{2(1-\cos k_y N_y)}{2M_\textrm{pl}^2 L_y^2}
    +E_\textrm{pl},
\end{align}
where $a$ is a lattice constant, $L_x(L_y)$ is the size of the magnetic unit cell of SkXs along the $x(y)$ axis, $N_x(N_y)$ is the number of sublattice sites within the magnetic unit cell along the $x(y)$ axis, and $(k_x,k_y)$ is defined within the 1st Brillouin zone of SkXs. Here, we have $N_x=27$ and $N_y=16$ in the skyrmion lattice, and set $L_x=100$~nm. 
Similarly, the magnon-plasmon coupling is approximated as
\begin{align}
H_\textrm{coupling}&\approx
\sum_{\vec{k}}
\sum_{j=1}^{N_\textrm{b}}
\textrm{i}G_{k_c}\left(\frac{\sin k_yN_y}{k_c L_y},-\,\frac{\sin k_xN_x}{k_c L_x},0\right)
\cdot
\left[
\vec{\mu}_j \,a_{\vec{k}}b_{-\vec{k},j}
+\vec{\mu}_j^* \, a_{\vec{k},\beta}b_{\vec{k},j}^\dagger
- \vec{\mu}_j \, a_{\vec{k}}^\dagger b_{\vec{k},j}
- \vec{\mu}_j^* \, a_{\vec{k}}^\dagger b_{-\vec{k},j}^\dagger
\right]
,\label{m-p_coupling2}
\end{align}
with the coupling strength is fixed at $G_{k_c}/h=0.98$~GHz.
As shown in Fig.~\ref{Fit_dispersion}(c), our fitting curve shows a good agreement with the dispersion of plasmons at $k=k_c$. 
Figure~\ref{Fit_dispersion}(d) shows the hybrid spectrum with the plasmon band replaced by the quadratic dispersion, where the Chern number below the hybridization gap is zero.
Since the CCW mode is characterized by $C=-1$ without magnon-plasmon coupling, it indicates the cancellation of the Chern number due to the magnon-plasmon coupling.

\begin{figure}
    \centering
    \includegraphics[width=90mm]{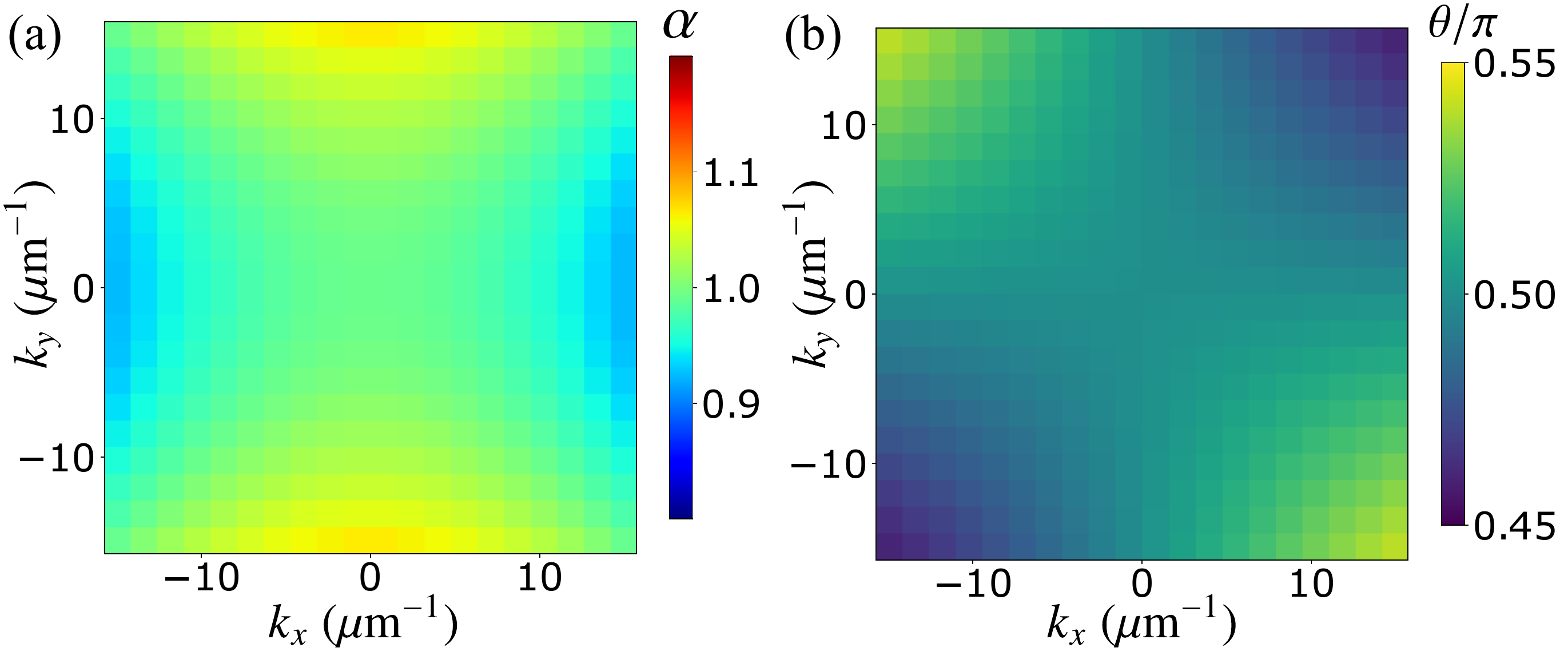}    
    \caption{
    (a)~The relative amplitude $\alpha_{\vec{k}}=|\mathcal{M}_{\vec{k},y}|/\mathcal{M}_{\vec{k},x}$ and (b) the complex phase $\theta_{\vec{k}}$ are plotted for the CCW mode.
    Parameters are defined as $D/J=0.5$, $B_zJ/D^2=0$, and $KJ/D^2=0.25$.
     }
    \label{DyM_1stBz}
\end{figure}

\subsection{Dynamical magnetization}

We consider the topological properties of magnon-plasmon hybrids using the dynamical magnetization~\cite{hirosawaMagnetoelectricCavityMagnonics2022}.
We introduce the Bogoliubov transformation for diagonalizing the spin-wave Hamiltonian:
\begin{align}
    \begin{pmatrix}
        b_{\vec{k},j} \\ b^\dagger_{-\vec{k},j}
    \end{pmatrix}
    =
    \sum_{n=1}^{N_\text{b}}
    T^{j,n}_{\vec{k}}
    \begin{pmatrix}
        b_{\vec{k},n} \\ b^\dagger_{-\vec{k},n}
    \end{pmatrix}=\begin{pmatrix}
         u^{j,n}_{\vec{k}} b_{\vec{k},n}+\left(v^{j,n}_{-\vec{k}} \right)^*b^\dagger_{-\vec{k},n} \\
         v^{j,n}_{\vec{k}}b_{\vec{k},n}+\left(u^{j,n}_{-\vec{k}}\right)^* b^\dagger_{-\vec{k},n}
    \end{pmatrix},
    \quad
    T^{j,n}_{\vec{k}}
    =
    \begin{pmatrix}
        u^{j,n}_{\vec{k}} & \left(v^{j,n}_{-\vec{k}} \right)^*\\
        v^{j,n}_{\vec{k}} & \left(u^{j,n}_{-\vec{k}}\right)^*
    \end{pmatrix},
    \label{eq:Bogo}
\end{align}
where $n$ is the band index. The magnon-plasmon coupling in Eq.~\eqref{m-p_coupling1} is rewritten as
\begin{align}
H_\textrm{coupling}&=
\sum_{n=1}^{N_\textrm{b}} 
\sum_{\vec{k}}
\textrm{i}G_k(\vec{\hat{k}}\times \hat{z})\cdot
\left[\vec{\mathcal{M}}_{\vec{k}}^{(n)\,*}\, b_{\vec{k},n}^\dagger a_{\vec{k}} - \vec{\mathcal{M}}_{\vec{k}}^{(n)}\, a_{\vec{k}}^\dagger b_{\vec{k},n}\right]
\nonumber\\
&=
\sum_{n=1}^{N_\textrm{b}} 
\sum_{\vec{k}}
\textrm{i}G_k
\left[\left(\hat{k}_y\mathcal{M}_{\vec{k},x}^{(n)\,*}-\hat{k}_x\mathcal{M}_{\vec{k},y}^{(n)\,*}\right)b_{\vec{k},n}^\dagger a_{\vec{k}} 
- \left(\hat{k}_y\mathcal{M}_{\vec{k},x}^{(n)}-\hat{k}_x\mathcal{M}_{\vec{k},y}^{(n)}\right)a_{\vec{k}}^\dagger b_{\vec{k},n}\right]
.
\end{align}
Anomalous terms are dropped in the above expression.
The dynamical magnetization of $n$-th magnon eigenstate is defined by
\begin{align}
\boldsymbol{\mathcal{M}}^{(n)}_{\vec{k}}
&= \sum_{ j=1 }^{N_\text{b}}\left(\vec{\mu}_{j} u^{j,n}_{\vec{k}} + \vec{\mu}^*_{j} v^{j,n}_{\vec{k}} \right).
\label{eq:MagnonM}
\end{align}
The relative phase difference between $\mathcal{M}_{\vec{k},x}$ and $\mathcal{M}_{\vec{k},y}$ determines the topology of the magnon-plasmon hybridization gap, although there is a degree of freedom for a global phase.
We fix $\mathcal{M}_{\vec{k},x}$ to be real and write $\mathcal{M}_{\vec{k},y}=\alpha_{\vec{k}} e^{\textrm{i}\theta_{\vec{k}}}$ where $\alpha_{\vec{k}}=|\mathcal{M}_{\vec{k},y}|/\mathcal{M}_{\vec{k},x}$ and $\theta_{\vec{k}}$ is the complex phase.
The magnon-plasmon coupling is then rewritten as
\begin{equation}
H_\textrm{coupling}=
\sum_{n=1}^{N_\textrm{b}} 
\sum_{\vec{k}}\textrm{i}G_k
\left[\mathcal{M}_{\vec{k},x}^{(n)}\left(\hat{k}_y-\hat{k}_x\alpha_{\vec{k}}^{(n)} e^{-\textrm{i}\theta^{(n)}_{\vec{k}}}\right)b_{\vec{k},n}^\dagger a_{\vec{k}} 
- \mathcal{M}_{\vec{k},x}^{(n)}\left(\hat{k}_y-\hat{k}_x\alpha^{(n)}_{\vec{k}} e^{\textrm{i}\theta^{(n)}_{\vec{k}}}\right)a_{\vec{k}}^\dagger b_{\vec{k},n}\right].
\end{equation}
We note that the complex phase $\theta_{\vec{k}}$ depends on the relative phase difference between time-dependent magnetization $M_x(t)$ and $M_y(t)$ when magnons are excited. Thus, magnons supporting rotating magnetization induce a topological hybridization gap with plasmons in general.

We consider the dynamical magnetization of the CCW mode. 
In Fig.~\ref{DyM_1stBz}(a,b), we plot $\alpha_{\vec{k}}$ and $\theta_{\vec{k}}$ of the CCW mode. 
Crucially, we find that $\alpha_{\vec{k}}=1$ and $\theta_{\vec{k}}=\pi/2$ near $\Gamma$ point, leading to the topological hybridization gap between magnons and plasmons.
The magnon-plasmon coupling of the CCW mode is obtained as
\begin{align}
H_\textrm{CCW}&=
-
\sum_{\vec{k}}
\textrm{i}G_k\mathcal{M}_{\vec{k},x}^{\textrm{CCW}}
\left[(\hat{k}_x-\textrm{i}\hat{k}_y)b_{\vec{k},n}^\dagger a_{\vec{k}} 
+(\hat{k}_x+\textrm{i}\hat{k}_y)a_{\vec{k}}^\dagger b_{\vec{k},n}\right]
\\
&
=
\sum_{\vec{k}}
G''_{\vec{k}}e^{-\textrm{i}\psi_k}b_{\vec{k},n}^\dagger a_{\vec{k}} 
+G''_{\vec{k}}e^{\textrm{i}\psi_k}a_{\vec{k}}^\dagger b_{\vec{k},n}
\label{CCW_Pl}
\end{align}
where $\psi_k=\tan\frac{\hat{k_y}}{\hat{k}_x}$ and $G''_{\vec{k}}=-\textrm{i}G_{k}\mathcal{M}_{\vec{k},x}^{\textrm{CCW}}$ is a real value.
We obtain $\mathcal{M}_{\vec{k},x}= 0.58$ at $\Gamma$ point, resulting in $|G''_{\vec{k}}/h|=0.55$~GHz. It agrees with the hybridization gap in Fig.~\ref{Fit_dispersion}(b).
We note that the phase factor in the magnon-plasmon coupling is complex conjugate to Eq.~\eqref{eq:HFM}.
Therefore, the Chern number of the lower hybrid band is +1. As a result, the total Chern number vanishes below the hybridization gap.

\begin{figure}
    \centering
    \includegraphics[width=120mm]{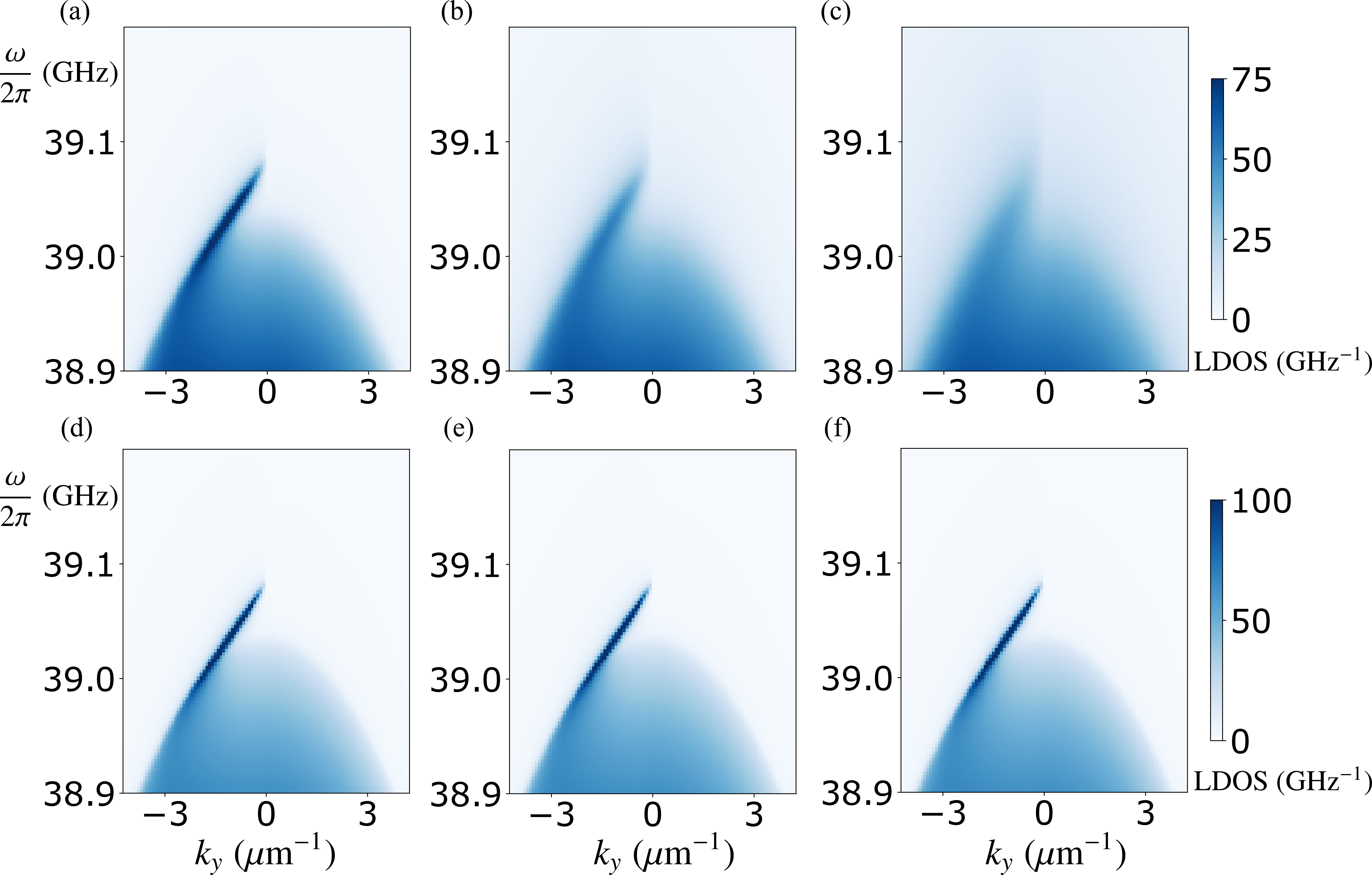}    
    \caption{
    (a-c) The LDOS at the edge of semi-infinite skyrmion lattices with increasing the damping rate of magnons as $\kappa_\textrm{mag}=8,20,40$~MHz, respectively. We fix the damping rate of plasmons at $\kappa_\textrm{pl}=4$~MHz. 
    (d-f) The LDOS at the edge of semi-infinite skyrmion lattices with increasing the damping rate of plasmons as $\kappa_\textrm{pl}=8,20,40$~MHz, respectively. We fix the damping rate of magnons at $\kappa_\textrm{mag}=4$~MHz. Parameters are provided in the Table~\ref{tab:parameters} and Sec.~\ref{sec:Energyscale_SkX}.
     }
    \label{damping_edge}
\end{figure}

\subsection{Topological magnon-plasmon hybrid}

We employ the renormalization approach to compute the local density of states~(LDOS) of a semi-infinite system~\cite{HENK199369, Mook2014edge}.
The magnetic textures of SkXs for semi-infinite calculations are prepared by simulating spin textures under the periodic boundary conditions.
We decompose the system into blocks of principal layers, in which only the nearest-neighbor layers interact.
The LDOS of magnons at the edge layer is given by
\begin{align}
    N_0(\omega,q_y)=-\frac{1}{\pi} \textrm{Im}\textrm{Tr}\Big[\Sigma G_{00}(\omega+\textrm{i}\kappa, q_y)\Big],
\end{align}
where $\kappa$ is the damping rate, and the subscript indicates the layer index.
The Green's function $G_{nm}$ satisfies
\begin{align}
    \delta_{nm}&=\sum_{j=0}^\infty ( z \Sigma \delta_{nj}-H^{(3)}_{nj})G_{jn}, 
\end{align}
with $z=\hbar(\omega+\textrm{i}\kappa)$ and $H_{nm}$ denotes the Hamiltonian matrix between $n$th and $m$th layers.
Here, we multiply $z$ by $\Sigma$ to account for the bosonic commutation relation.   
Iterative calculations lead to the vanishing of off-diagonal Hamiltonian matrices ($|H_{nm}|\rightarrow 0$), allowing us to obtain the edge Green's function $G_{00}$.

First, we set the damping constant $\kappa_\textrm{pl}=\kappa_\textrm{mag}=4$~MHz. Figure~4(b) of the main text shows the LDOS at the edge of semi-infinite skyrmion lattices over the 1st Brillouin zone.
The magnonic chiral edge state is obtained at $k_y<-10~\mu\textrm{m}^{-1}$ and ranges from 37~GHz to 60~GHz. 
We find the left-moving edge state at $x=1$ while the right-moving edge state should appear at $x\rightarrow \infty$.
In contrast, we expect the magnon-plasmon hybrid edge mode inside the small hybridization gap of 1.1~GHz at $k_y=k_c$.
Thus, these two oppositely propagating edge states exhibit a significant mismatch in energy and wave vector. 
Figure~4(c) of the main text shows the LDOS over a small window of energy and wave vector.
The magnon-plasmon edge state is obtained at $|k_y|<3~\mu\textrm{m}^{-1}$, connecting the almost flat magnon band and the plasmon band.

We also investigate the robustness of the magnon-plasmon edge state with larger damping rates.
Figure~\ref{damping_edge}(a-c) shows the LDOS at the edge of semi-infinite skyrmion lattices when the magnon damping rate is increased as $\kappa_\textrm{mag}=8,20,40$~MHz, respectively.
Although an increase in the magnon damping rate leads to the broadening of the LDOS, the magnon-plasmon edge state is still observed at $\kappa_\textrm{mag}=40$~MHz.
In comparison, the damping rate of plasmons does not affect the LDOS of the magnon-plasmon edge state even at $\kappa_\textrm{pl}=40$~MHz, as shown in Fig.~\ref{damping_edge}(d-f).
This finding indicates that the magnon-plasmon edge state is dominantly occupied by magnons due to the significant difference in their group velocities.